%% file: main.tex
\documentclass[twocolumn,tighten,table]{aastex62}
\pdfoutput=1 

\usepackage{amsmath,amstext}
\usepackage[T1]{fontenc}
\usepackage{apjfonts}
\usepackage[figure,figure*]{hypcap}
\usepackage{url}
\input{hyperlink-year-only-natbib-patch}

\newcommand*{\msun}{\text{M}_\odot}
\newcommand*{\sfr}{\rm{SFR}}
\newcommand*{\mstar}{M_{\star}}
\newcommand*{\magr}{M_{\rm r}}
\newcommand*{\mhalo}{M_{\rm halo}}
\newcommand*{\mpeak}{M_{\rm peak}}
\newcommand*{\dndmpeak}{{\rm dn/d{\it M}_{peak}}}
\newcommand*{\dndmpeakfit}{{\rm dn/d{\it M}_{peak}^{fit}}}
\newcommand*{\mpeaktwo}[1]{M_{\rm peak}^{\rm #1}}
\newcommand*{\HI}{H\,{\sc i}}
\newcommand*{\HII}{H\,{\sc ii}}
\newcommand*{\HeII}{He\,{\sc ii}}
\newcommand*{\OII}{O\,{\sc ii}}
\newcommand*{\OIII}{O\,{\sc iii}}
\newcommand*{\NII}{N\,{\sc ii}}
\newcommand*{\SII}{S\,{\sc ii}}

\definecolor{alexiegreen}{HTML}{0CA82B}

\shorttitle{CosmoDC2}
\shortauthors{Korytov et al.\ (LSST~DESC)}

\begin{document}
\title{CosmoDC2: A Synthetic Sky Catalog for Dark Energy Science with LSST}

\author{Danila Korytov}
\affiliation{Argonne National Laboratory, Lemont, IL 60439, USA}
\affiliation{Department of Physics, University of Chicago, Chicago, IL 60637, USA}

\author{Andrew Hearin}
\affiliation{Argonne National Laboratory, Lemont, IL 60439, USA}

\author[0000-0002-2545-1989]{Eve Kovacs}
\affiliation{Argonne National Laboratory, Lemont, IL 60439, USA}

\author{Patricia Larsen}
\affiliation{Argonne National Laboratory, Lemont, IL 60439, USA}

\author{Esteban Rangel}
\affiliation{Argonne National Laboratory, Lemont, IL 60439, USA}

\author[0000-0002-8658-1672]{Joseph Hollowed}
\affiliation{Argonne National Laboratory, Lemont, IL 60439, USA}

\author[0000-0001-5501-6008]{Andrew~J.~Benson}
\affiliation{Carnegie Observatories, Pasadena, CA 91101, USA}

\author[0000-0003-1468-8232]{Katrin~Heitmann}
\affiliation{Argonne National Laboratory, Lemont, IL 60439, USA}

\author[0000-0002-1200-0820]{Yao-Yuan~Mao}
\affiliation{Department of Physics and Astronomy, \& Pittsburgh Particle Physics, Astrophysics and Cosmology Center (PITT PACC),\\ University of Pittsburgh, Pittsburgh, PA 15260, USA}

\author{Anita Bahmanyar}
\affiliation{Astronomy and Astrophysics, University of Toronto, Toronto, Ontario M5S 1A1, Canada}

\author{Chihway Chang}
\affiliation{Department of Physics, University of Chicago, Chicago, IL 60637, USA}

\author{Duncan~Campbell}
\affiliation{McWilliams Center for Cosmology and Department of Physics, Carnegie Mellon University, Pittsburgh, PA 15213, USA}

\author{Joseph~DeRose}
\affiliation{Kavli Institute for Particle Astrophysics and Cosmology \& Department of Physics, Stanford University, Stanford, CA 94305, USA}

\author{Hal Finkel}
\affiliation{Argonne National Laboratory, Lemont, IL 60439, USA}

\author{Nicholas Frontiere}
\affiliation{Argonne National Laboratory, Lemont, IL 60439, USA}
\affiliation{Department of Physics, University of Chicago, Chicago, IL 60637, USA}

\author[0000-0003-1530-8713]{Eric~Gawiser}
\affiliation{Department of Physics and Astronomy, Rutgers University, Piscataway, NJ 08854, USA}

\author[0000-0002-7832-0771]{Salman~Habib}
\affiliation{Argonne National Laboratory, Lemont, IL 60439, USA}

\author{Benjamin Joachimi}
\affiliation{Department of Physics and Astronomy
University College London, London, WC1E 6BT, UK}

\author[0000-0001-7956-0542]{Fran\c{c}ois~Lanusse}
\affiliation{McWilliams Center for Cosmology and Department of Physics, Carnegie Mellon University, Pittsburgh, PA 15213, USA}

\author[0000-0001-6800-7389]{Nan Li}
\affiliation{School of Physics and Astronomy, University of Nottingham, University
Park, Nottingham, NG7 2RD, UK}

\author[0000-0003-2271-1527]{Rachel~Mandelbaum}
\affiliation{McWilliams Center for Cosmology and Department of Physics, Carnegie Mellon University, Pittsburgh, PA 15213, USA}

\author{Christopher Morrison}
\affiliation{Department of Astronomy, University of Washington, Seattle, WA 98105, USA}

\author[0000-0001-8684-2222]{Jeffrey~A.~Newman}
\affiliation{Department of Physics and Astronomy, \& Pittsburgh Particle Physics, Astrophysics and Cosmology Center (PITT PACC),\\ University of Pittsburgh, Pittsburgh, PA 15260, USA}

\author[0000-0003-2265-5262]{Adrian~Pope}
\affiliation{Argonne National Laboratory, Lemont, IL 60439, USA}

\author{Eli Rykoff}
\affiliation{SLAC National Accelerator Laboratory, Menlo Park, CA, 94025, USA}

\author{Melanie~Simet}
\affiliation{Jet Propulsion Laboratory, California Institute of Technology, Pasadena, CA 91109, USA}

\author{Chun-Hao To}
\affiliation{Kavli Institute for Particle Astrophysics and Cosmology \& Department of Physics, Stanford University, Stanford, CA 94305, USA}

\author{Vinu~Vikraman}
\affiliation{Argonne National Laboratory, Lemont, IL 60439, USA}

\author{Risa H. Wechsler}
\affiliation{Kavli Institute for Particle Astrophysics and Cosmology \& Department of Physics, Stanford University, Stanford, CA 94305, USA}

\author{Martin White}
\affiliation{Department of Physics, University of California, Berkeley, 366 LeConte Hall MC 7300, Berkeley, CA 94720, USA}

\collaboration{(The LSST Dark Energy Science Collaboration)}

\begin{abstract}
This paper introduces cosmoDC2, a large synthetic galaxy catalog designed to support precision dark energy science with the Large Synoptic Survey Telescope (LSST). CosmoDC2 is the starting point for the second data challenge (DC2) carried out by the LSST Dark Energy Science Collaboration (LSST DESC).  The catalog is based on a trillion-particle, ($4.225$~Gpc)$^3$ box cosmological N-body simulation, the `Outer Rim' run. It covers $440$~deg$^2$ of sky area to a redshift of $z=3$ and is complete to a magnitude depth of 28 in the $r$-band. Each galaxy is characterized by a multitude of properties including stellar mass, morphology, spectral energy distributions, broadband filter magnitudes, host halo information and weak lensing shear. The size and complexity of cosmoDC2 requires an efficient catalog generation methodology; our approach is based on a new hybrid technique that combines data-driven empirical approaches with semi-analytic galaxy modeling.  A wide range of observation-based validation tests has been implemented to ensure that cosmoDC2 enables the science goals of the planned LSST~DESC DC2 analyses. This paper also represents the official release of the cosmoDC2 data set, including an efficient reader that facilitates interaction with the data.
\end{abstract}


%\accepted{June ??, 2019}
\submitjournal{the Astrophysical Journal Supplement}

\section{Introduction}
\label{sec:intro}
The next generation of
large imaging and spectroscopic survey projects including LSST \citep{lsst}, the Wide Field Infrared Survey Telescope (WFIRST) \citep{wfirst}, the Dark Energy Spectroscopic Instrument (DESI) \citep{2016arXiv161100036D} and Euclid \citep{euclid} will provide a wealth of data for modern cosmological analyses that seek to probe the nature of dark energy. The advent of these large data sets heralds a new era in cosmology that is characterized by  small statistical uncertainties, improved control of systematic errors, and unique opportunities to combine multiple probes of dark energy, all leading to much improved constraints on this little understood component of the Universe. The LSST Dark Energy Science Collaboration (LSST~DESC)~\citep{2012arXiv1211.0310L} has convened to prepare for the large and complex data set that will arrive with the commencement of LSST operations in 2022. Since the data will become publicly available almost immediately, scientists must have robust, well-understood analysis pipelines in place. This goal would not be achievable without extensive simulation campaigns aimed at providing simulated skies in the form of synthetic galaxy catalogs.

These galaxy catalogs play a number of essential roles in cosmological surveys. They serve as testbeds to study questions of survey design, to enable studies of possible systematics (e.g., fiber collisions, telescope point-spread functions, shape measurement assumptions), to facilitate tests of data reduction pipelines and to provide reference data that can be used to validate analysis pipelines for cosmological inference. The demands of contemporary surveys for high-quality synthetic skies are only becoming more stringent as the field progresses further into an era in which cosmological inference is systematics-limited.

Given sufficient understanding of the underlying physics, such catalogs can be thought of as the result of forward models that provide approximate theoretical predictions for the galaxies inhabiting a physical universe. As the models improve over time, the predictions become ever more faithful to the properties of the observed Universe. Consequently, synthetic catalogs, and summary statistics derived from them, enable cosmological inference and function as tests of robustness for these methods.

 A wide variety of techniques is currently being used to construct large-volume synthetic galaxy catalogs for contemporary and near-future galaxy surveys \citep[see][for an overview]{wechsler_tinker}. At the extreme end of computational expense, hydrodynamical simulations are the closest approximation in the field to an {\em ab initio} model of galaxy formation.
These simulations track the evolution of the dark matter and the baryons under the influence of gravity and gas physics and include sub-grid models for a number of astrophysical processes such as gas cooling and heating, star formation, and supernova and AGN feedback. Contemporary approaches include simulated cosmological volumes of the scale of $\mathcal{O}(100)$ Mpc \citep{khandai_etal14, schaye_etal15, springel_etal18} to $\mathcal{O}(1)$ Gpc \citep{dolag_etal15, emberson_etal18} as well as zoom-in simulations at much higher resolution \citep{agertz_etal12, brooks_zolotov12, agora, nihao, hopkins_etal18}.
While hydrodynamical simulations are indispensable to the study of systematic effects associated with baryonic physics, these simulations are not typically used to produce large-scale galaxy catalogs for sky surveys due to their computational expense, which arises from the necessity of solving for gas dynamics at high spatial and mass resolution, and resolving small spatio-temporal scales when including sub-grid models. Additionally, parameterized sub-grid models are typically integrated within the main simulation and cannot be conveniently ``bolted on'' after the simulation is completed. This significantly increases the number of simulations needed to explore the associated parameter space and to calibrate the models against observations. 

A less computationally expensive approach involves the use of
semi-analytic models of galaxy formation (SAMs). In these models,
the synthetic catalog is generated using gravity-only simulations coupled with additional modeling for the baryonic physics not contained in the underlying simulation. This additional modeling is based on the mass-assembly history of each halo \citep{kauffmann99,croton06,benson_2010b,lacey16}. Hence, all SAMs are predicated upon connecting the properties of individual galaxies with individual dark matter halos. The SAM approach is to parameterize baryon-specific processes as functions of the halos and their evolution; on a halo-by-halo basis SAMs seek to model directly how baryons would have evolved had they been included in the N-body simulation \citep[see][for a recent review]{somerville_dave15}. Using SAMs to generate synthetic catalogs for large surveys
is still computationally demanding, particularly for imaging surveys in need of large quantities of synthetic data for faint galaxies. Furthermore, while considerable recent progress has been made in our ability to explore the parameter space of contemporary SAMs \citep{henriques_etal15,vandaalen16,ruiz15}, calibrating a SAM to high accuracy is nonetheless computationally expensive.

When generating synthetic data for contemporary galaxy surveys, such difficulties are compounded by the evolving nature of the validation criteria used to evaluate the fitness of the catalog. As additional scientists join and become active in a collaboration, their expertise informs changes or refinements to the criteria; adjustments may also be made as surveys release new data. The development of new analysis techniques may also require synthetic data to have properties in addition to those originally planned. This evolving nature is fundamental to the operating mode of collaborations conducting large galaxy surveys. As a consequence, when generating simulated catalogs to complement a survey, the underlying model should be straightforward to calibrate, since in practice the parameters may need to be refit many times over the course of the survey.

To address the aforementioned challenges associated with using SAMs, many collaborations have instead used empirical models to generate their synthetic data.
In this approach, one assumes the existence of simple scaling relations between dark matter halos and galaxies, and fits the parameters of these scaling relations to observational data. Empirical models are computationally more efficient than SAMs; therefore, the parameters of these models are less expensive to fit. When generating synthetic catalogs with empirical methods the challenge lies in attaining sufficient realism to reflect the level of complexity needed to satisfy survey requirements. For example, in the MICE catalogs~\citep{2015MNRAS.447.1319F} created for the Dark Energy Survey (DES), the authors begin with a standard form of the Halo Occupation Distribution \citep[HOD, ][]{berlind03, Kravtsov2004, zheng07} to inject synthetic galaxies into simulated host halos. After tuning their baseline HOD model, the properties of galaxies they augmented with additional attributes, such as broadband color, using data-driven methods. A similar methodology is used to generate the Euclid Flagship simulation galaxy catalog\footnote{\url{http://sci.esa.int/euclid/59348-euclid-flagship-mock-galaxy-catalogue}} based on a large gravity-only simulation described in~\citet{2017ComAC...4....2P}. Similarly, the Buzzard catalogs developed for DES \citep{buzzard_flock} use a baseline abundance-matching approach  \citep{Kravtsov2004,Conroy2006,behroozi10,moster10,Reddick2013} on a high-resolution simulation to generate a tuning catalog. The ADDGALS technique~(R. Wechsler et al. 2019, in preparation) is used to populate lower-resolution simulations with galaxies based on this tuning catalog. The galaxies are then assigned template spectral energy distributions (SEDs) using empirical techniques.

This paper presents the cosmoDC2 synthetic galaxy catalog. It is one of the most ambitious synthetic catalogs ever constructed for survey science, as well as one of the most complex, containing highly nonlinear correlations between a large number of galaxy properties spanning a multidimensional parameter space. Our approach combines empirical methods with semi-analytic modeling using the Galacticus code~\citep{benson_2010b} to produce a catalog which is easily calibrated while maintaining a high degree of physical realism. The input to cosmoDC2 is a large-volume, N-body (gravity-only) cosmological simulation, the Outer Rim run, carried out using the Hybrid/Hardware Accelerated Cosmology Code (HACC)~\citep{2016NewA...42...49H}. For the construction of the catalog, data products from the smaller AlphaQ simulation, and the UniverseMachine~\citep{behroozi_etal18} modelling approach  
are used in the semi-analytic and empirical modeling components, respectively.

CosmoDC2 was designed for the second LSST DESC Data Challenge (DC2), an ambitious effort to generate a data set that is very similar to a multi-year LSST data release. DC2 not only aims to improve analysis pipelines in advance of the arrival of LSST imaging, but also serves the important function of uniting hundreds of LSST~DESC members with a specific and concrete data set to analyze with common software tools.
The design of cosmoDC2 was driven by a number of science goals that are applicable to any synthetic catalog intended to provide simulated data for a wide and deep sky survey such as LSST. These goals are quantified by specific criteria applied to a large number of tests that have been developed by the LSST~DESC analysis working groups and incorporated into DESCQA \citep{descqa}, the LSST~DESC automated validation framework for synthetic sky catalogs. These tests compare the catalog contents with judiciously chosen observational data, enabling an assessment of the synthetic catalog's fidelity to the statistics of the real sky. The validated catalog can then be used to provide realistic inputs for scientific analyses as described above and to serve as the input for extragalactic component of DC2 image simulations.

Broadly speaking, the need to test analysis pipelines for probes involving static measurements such as weak lensing, large-scale structure, galaxy clusters, and photometric redshifts leads to a requirement that the full catalog cover an area of thousands of square degrees, which in turn demands that a very large input simulation underlies the catalog. However the area requirement for the image simulations is relatively modest in comparison. The catalog is, therefore, being released in 2 stages: the first stage provides enough area (440 deg$^2$) to run the DC2 image simulations; the second stage, to be completed in the near future, will provide the larger area (5000 deg$^2$) needed to enable analysis tests requiring high statistics.
The complexity of the validation tests also leads to stringent requirements on the properties of the galaxy distribution and their connection to the underlying dark matter distribution. In particular, for weak lensing quantities to be correctly estimated at each galaxy position a full ray-tracing code based on the Outer Rim particle lightcones is needed.
As described below, all of these considerations informed the methodology used to produce the cosmoDC2 catalog.

The rest of the paper is organized as follows. 
In Sec.~\ref{sec:e-to-e} we provide a general overview of the end-to-end workflow for generating the synthetic sky catalog. We briefly describe the gravity-only simulations and related data products underlying the catalog (halos, merger trees and lightcones) in Sec.~\ref{sec:sim}. Sec.~\ref{sec:shear} describes the generation of the weak lensing catalog. Sec.~\ref{sec:gc} discusses the galaxy catalog, where we give a brief summary of the catalog generation and the catalog content. The release of this paper is coincident with the public release of the 440 sq. deg. cosmoDC2 catalog. Selected results from the catalog-validation tests are shown in Sec.~\ref{sec:results}. A concluding discussion and future outlook is provided in Sec.~\ref{sec:outlook}.

\section{CosmoDC2 Production Overview}
\label{sec:e-to-e}

\begin{figure}
\centering
\includegraphics[width=9cm]{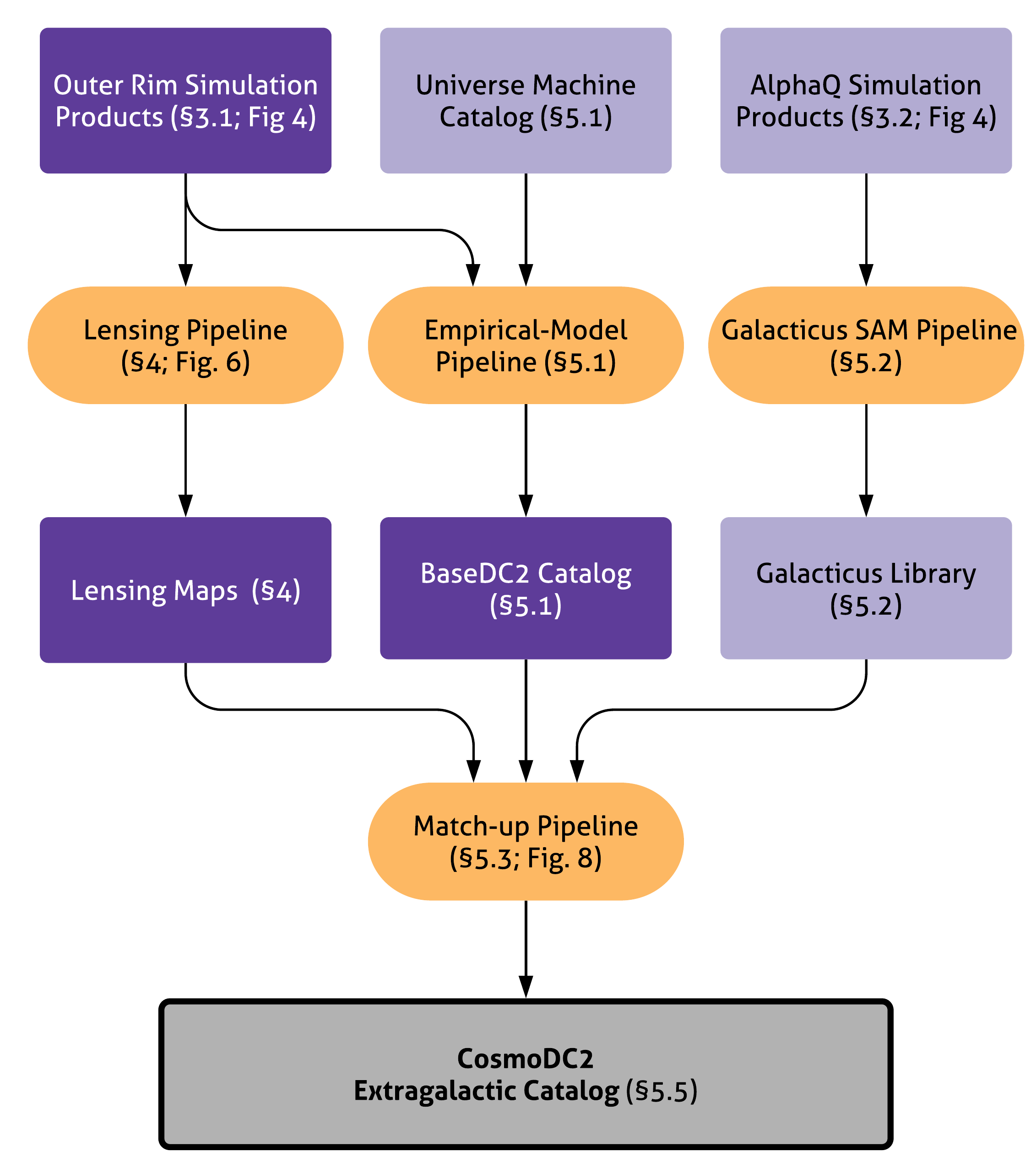}
\caption{Conceptual overview of the workflow to produce cosmoDC2. Data products are shown as rectangles in dark and light purple for data derived from the Outer Rim and the smaller auxiliary simulations, respectively. Pipelines are shown as ovals in light orange.
Numbers in parentheses refer to the sections and figures in this work where a detailed descriptions are given.
}
\label{overview}
\label{fig:overview}
\end{figure}

The key aspects of our production pipeline are shown in Figure~\ref{overview}, which provides a conceptual overview of how we use the simulation data products available from both the large Outer Rim simulation
and the smaller UniverseMachine and AlphaQ simulations 
to produce the cosmoDC2 extragalactic catalog. As noted in the figure, the data products and pipelines are described in detail in subsequent sections of the paper. The first stage of the workflow shows how the data products from each simulation are used as inputs to separate pipelines
to produce corresponding intermediate data products. Next, these intermediate data products are combined in a final pipeline to produce the cosmoDC2 extragalactic catalog.

The Lensing Pipeline (Sec.~\ref{sec:shear}) inputs particle data from the Outer Rim simulation to a lightcone generator, projects these particles onto discrete redshift shells and computes the surface density using a density estimator, and traces photons through these shells using a ray-tracing code, to generate curved-sky lensing maps. These lensing maps include the deflections experienced by photons as well as the image distortions in the weak lensing regime.

The Empirical Model Pipeline (Sec.~\ref{subsec:empirical}) uses halo and merger-tree data from the Outer Rim simulation combined with Monte Carlo resampling of galaxies from the UniverseMachine simulation (version ERD) to produce the baseDC2 catalog. BaseDC2 contains all of the cosmoDC2 galaxies but with a limited set of properties. The UniverseMachine simulation  consists of galaxies that have been assigned to halos in the MultiDark Planck 2 (MDPL2) N-body simulation \citep{klypin_etal16} using  the UniverseMachine prescription described in \citet{behroozi_etal18}. The numbers, the positions and properties (stellar mass, $\mstar$, and star-formation rate, \sfr) for these galaxies are determined by tuning the UniverseMachine-model parameters to observational data. The tuning procedure takes into account errors in the observational data by marginalizing over nuisance parameters that represent the additional sources of systematic errors required to achieve global fits to multiple observational data sets. We use further empirical modeling to augment the properties of these galaxies to include a limited selection of LSST rest-frame colors and magnitudes. These properties are assigned probabilistically.

The SAM pipeline (Sec.~\ref{subsec:sam}) uses merger trees built from the small AlphaQ simulation
as inputs to the Galacticus SAM \citep{benson_2010b} to produce the Galacticus library of galaxies. These galaxies have an extensive set of self-consistent properties that have been obtained by solving a set of ordinary differential equations (ODEs) following the mass assembly history of each galaxy's host halo. Galacticus nonetheless has simulation-dependent model parameters that must be tuned in order to produce results that match observational data. Since this tuning was beyond the scope of our work, the statistical ensemble of properties for the galaxies modeled by Galacticus does not fully meet our strict validation requirements. In our hybrid approach, however, we can use this set of galaxies as a library from which to select appropriate galaxies with a complex set of properties. The additional properties obtained from the Galacticus library are required by the validation criteria that have been supplied by the DESC science working groups.

The selection of library galaxies is performed in the Match-up Pipeline which uses the three intermediate data products as inputs. The lensing maps are used to interpolate the values of the weak lensing quantities (shear, convergence and deflection angles) to the position of each baseDC2 galaxy. Then the pipeline finds a suitable matching galaxy from the Galacticus Library for each galaxy in the baseDC2 catalog and augments the properties of the baseDC2 galaxies with those of the library galaxy, thus adding realistic complexity to the galaxy model. Additional empirical modeling is done after the match-up.

The workflow shows how our hybrid method combining empirical modeling and SAMs
leverages the data from both small and large simulations to produce a realistically complex synthetic catalog. It is important to note that in order to tune the properties of the galaxies to better match any validation data, it is only necessary to change the empirical model and rerun the Match-up Pipeline, both of which are relatively inexpensive. Since we are using the results from the Galacticus SAM as a library, that computationally expensive step in the workflow need only be carried out once. It is much less expensive to iterate over parameters in the empirical model than to tune the parameters for the SAM. Our workflow makes it feasible to iterate the galaxy properties multiple times to achieve good agreement with the validation data.  Critically, the distributions of galaxy properties of the SAM galaxies {\it span} the ranges of observed values. If this is not the case, the matching procedure will fail in some regions and some additional strategies will be required to generate a good match. (See Sec.~\ref{subsec:match}.)

For the program outlined above to be successful, it is important that the empirical model has the flexibility to produce an accurate realization of the limited galaxy properties, and that these properties are sufficiently correlated with other galaxy properties of interest. In our case, the UniverseMachine model generates stellar mass ($\mstar$) and star-formation rate ($\sfr$) distributions that match observational data. Additional empirical modeling generates selected colors and luminosities that are based on these $\mstar$ and $\sfr$ values and exploits the inherent correlations between stellar mass, star-formation rates and the brightness and color of galaxies. Matching to a SAM-library galaxy to obtain a complex set of properties assumes that the stellar mass and selected brightness and colors are sufficiently correlated with the full set of galaxy properties to obtain realistic distributions of the latter.  If all of these assumptions are fulfilled, then the combination of empirical modeling with SAM resampling should generate a catalog of galaxies with a complex set of properties whose summary statistics satisfy the validation criteria.

\section{The Underlying Simulations}
\label{sec:sim}

Two gravity-only simulations underlie the construction of the cosmoDC2 catalog. The Outer Rim simulation serves as the basis of the final synthetic galaxy catalog including weak lensing quantities. A smaller simulation, called AlphaQ, was used for generating the Galacticus-based galaxy library. We now provide a description of both simulations as well as the related data products, such as halo catalogs, merger trees, and lightcones.

\subsection{The Outer Rim Simulation}

\label{sec:sim_or}

The Outer Rim simulation (for details and results on basic statistics such as power spectra and mass functions at different redshifts, see \citealt{heitmann2019}) is one of the largest simulations at its mass resolution available world-wide and was carried out with the Hardware/Hybrid Accelerated Cosmology Code (HACC)~\citep{2016NewA...42...49H}. HACC has been developed to run on all currently available and planned supercomputer architectures, including many-core machines and GPU-accelerated systems. The Outer Rim run was carried out on 32 racks of the IBM BG/Q system Mira (two-thirds of the entire machine) at the Argonne Leadership Computing Facility (ALCF). On many-core systems, such as the BG/Q, HACC uses a high-order spectral particle mesh method, combined with a tree calculation for the short-range forces~\citep{2016NewA...42...49H}. For the analysis of the simulation we employed Mira itself as well as Cooley, a powerful GPU-enhanced analysis cluster at the ALCF.

The Outer Rim simulation covers a volume of (4.225 Gpc)$^3$ sampled with 10,240$^3$ tracer particles, leading to a mass resolution of $m_p=2.6\cdot 10^9 $M$_\odot$. We used a cosmological model close to the best-fit WMAP-7 parameter set~\citep{Komatsu11} given by
$\omega_{\rm cdm}=0.1109$, $\omega_{\rm b}=0.02258$, $n_s=0.963$, $h=0.71$, $\sigma_8=0.8$, and $w=-1.0$.

We saved 101 time snapshots from $z\sim 10$ down to $z=0$, evenly spaced in $\log_{10}(a)$, where $a$ is the scale factor. (Two snapshots were corrupted on disk before we were able to fully analyze them, leading to a final number of 99 snapshots). The spacing between snapshots varies from $\sim 300$ Myr for snapshots with $z \sim 0$, $\sim 200$ Myr for snapshots with $z \sim 1$, and $\sim 80$ Myr for snapshots with $z \sim 3$. For each snapshot we saved the full particle output and 1\% randomly downsampled particles. The full particle outputs are stored on tape, while the downsampled outputs are kept on disk for easy accessibility. An extensive analysis for each snapshot was carried out as detailed in \cite{heitmann2019}. The analysis steps relevant for the present work are summarized in Sec.~\ref{sec:halos}.

\subsection{The AlphaQ Simulation}

The AlphaQ simulation covers a simulation volume of (360.56 Mpc)$^3$ and evolves 1024$^3$ particles. This yields a mass resolution of $m_p=1.6\cdot 10^9$M$_\odot$, close to that of the Outer Rim simulation, but with a volume that is approximately 1600 times smaller. We carry out exactly the same analyses as for the Outer Rim run, and save data products at the same time steps. Initially, the AlphaQ simulation was used for prototyping the workflow to generate the synthetic galaxy catalog in order to avoid having to handle large amounts of data from the very start. This resulted in the `protoDC2' catalog, which was used for DC2 pipeline development. The AlphaQ simulation is also the source of the Galacticus-generated galaxy library that is used to assign complex galaxy properties to the final cosmoDC2 catalog.

\subsection{Halo Catalogs and Merger Trees}
\label{sec:halos}

The halo catalogs are generated using a parallel Friends-of-Friends (FOF)-based halo finder with a linking length of $b=0.168$ and a minimum requirement of 20 particles per halo.
Halo merger trees (based on the FOF catalogs) are constructed using a newly developed particle-membership algorithm~\citep{rangel2017building} aiming to address the inconsistencies that arise from temporary mergers. Working in reverse sequential order, pairs of temporally adjacent snapshots are processed to identify halo progenitor/descendant relationships, simultaneously replacing halos that have split, i.e., halos with multiple descendants, with their individual ``fragment'' components. In this way, the fragment halos propagate the splitting event through the analysis and ensure at most a single merging event for every halo.

\subsection{Lightcone Generation}
\label{sec:lightcone}

We build particle and halo lightcones by tiling the Outer Rim simulation box in space, and applying a parallel solver which interpolates objects between adjacent snapshot positions to find their location of null spacetime separation from an observer at the origin. The cosmoDC2 observed density field is created by building a particle lightcone from Outer Rim snapshots randomly downsampled to 1\%, through which lensing observables are simulated (see Sec.~\ref{sec:shear}). We also construct an accompanying halo lightcone built from the simulation's FOF merger tree, upon which galaxies are later placed (see Sec.~\ref{subsubsec:restframe_colors}). Each of these lightcones fills one octant (\textasciitilde 5,000 deg$^2$) of the sky, and has a depth of $z=3$. A high-level description of the particle and halo lightcone generation is given below in Sec. ~\ref{sec:plc} and ~\ref{sec:hlc}, respectively, and technical details of the solver implementation can be found in Appendix ~\ref{sec:lc_appdx}.

\subsubsection{Particle Lightcone}
\label{sec:plc}

A lightcone, as built from N-body simulations, can be thought of as a set of concentric shells centered on the observer, where the boundaries of those shells are determined by the discrete redshifts of each simulation output snapshot. In filling the lightcone, we require a general prescription to solve for the contents of each shell. Numerous such methods are described in the literature; previous efforts have chosen to fill the lightcone volume with simulation objects which retain their snapshot positions and velocities \citep{Blaizot:2003av, Kitzbichler:2006ec}, while others implement linear and higher order schemes to interpolate particle positions between snapshots \citep{evrard2002, 2013MNRAS.429..556M, Smith:2017tzz}.

The optimal choice between these approaches depends on the characteristics of the underlying simulation; if the time resolution of the outputs is coarse, then interpolation may cause excessive smoothing of the density field on small scales. On the other hand, simply concatenating shells extracted from snapshots introduces discontinuities in the correlation function across redshift. Given Outer Rim's relatively high spatial and temporal resolution (57 snapshot outputs to z=3), we find that a linear interpolation method (acceleration and higher order position derivatives are assumed to be zero; see Appendix~\ref{sec:lc_appdx} for details) is sufficient.

\begin{figure}[h!]
  \centering
  \includegraphics[width=\linewidth]{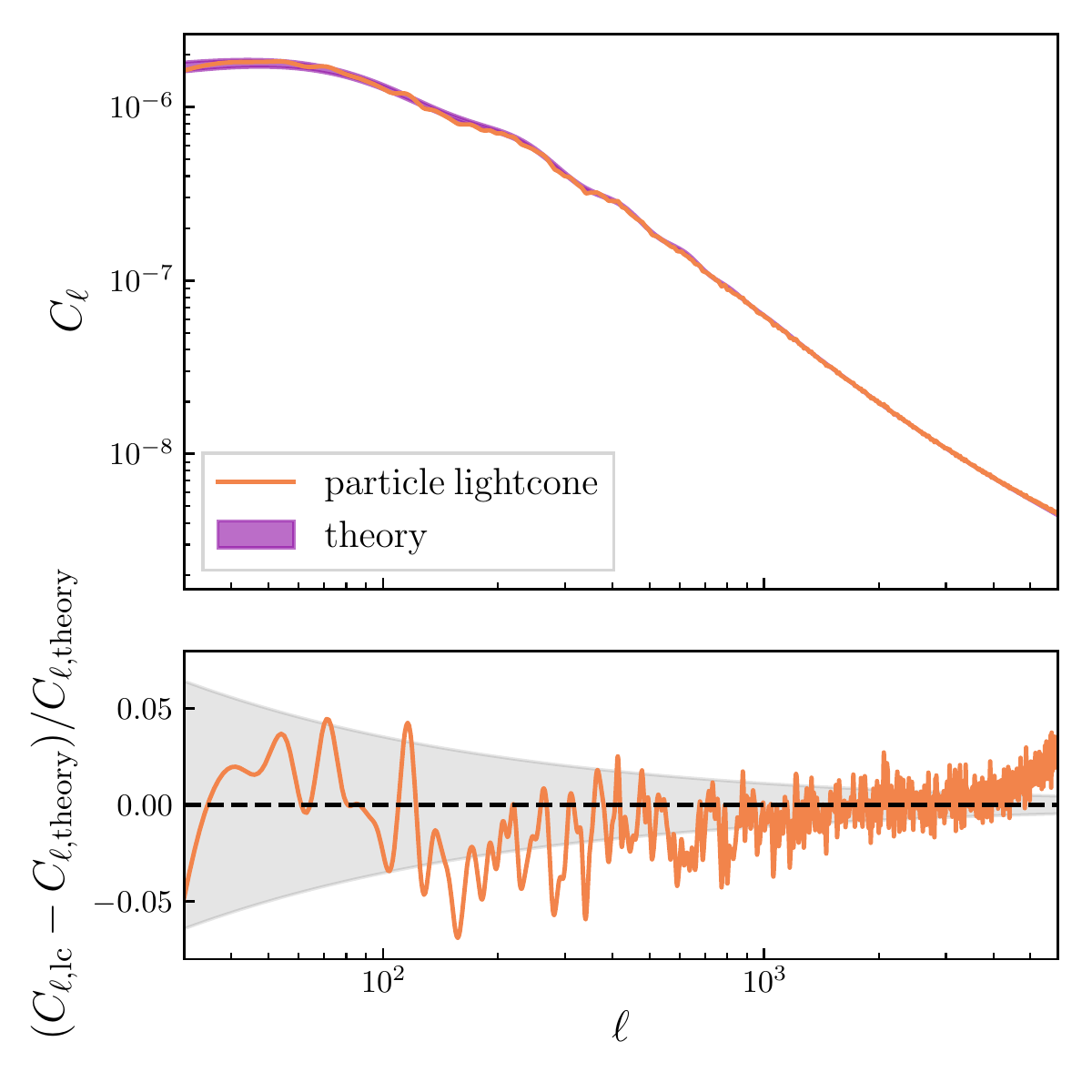}

\caption{ Projected particle overdensity power spectrum (top) and residuals with respect to theory (bottom) for a shell of width of approximately $ 450\rm{Mpc}$
at $z\approx 1.7$, after shot-noise corrections. The theory curve is obtained from the CosmicEmu power spectrum emulator, and has been corrected for expected levels of smoothing due to finite shell-width effects. The shaded region in the residual plot shows the approximate $1\sigma$ level of cosmic variance given the Polspice kernels. The up-turn in the residuals at small scales is due to the interpolation errors; these are sub-percent below $\ell\approx3000$, rising to approximately $3\%$ by $\ell=6000$.
}
\label{fig:lc_pk}
\end{figure}

The angular overdensity power spectrum of a section of the particle lightcone at $z\approx 1.7$ resulting from our procedure being applied to downsampled Outer Rim snapshots is shown in Figure~\ref{fig:lc_pk}. This is computed using the Polspice code \citep{polspice}, and corrected for shot noise. We compare this to a theory prediction based on the CosmicEmu power spectrum~\citep{2017ApJ...847...50L} corrected for finite shell-width effects
\citep[see, e.g., ][for a description of these effects]{2017ApJ...850...24T}. At large scales there is good agreement with the model to within the cosmic variance and expected levels of model inaccuracy. On small scales where we expect the particle interpolation to increase the power (see, e.g., \citealt{2013MNRAS.429..556M}), we see biases of \textasciitilde$1-3\%$.

The extremely large volume that will be probed by LSST poses a challenge when solving for an observer's past lightcone. In particular, the DC2 catalog effort needs to model galaxies out to redshift $z\lesssim3,$ where the comoving distance is in excess of 6~Gpc, while Outer Rim extends only to 4.225~Gpc in each dimension. Our strategy, then, is to build a lattice of replicated simulation volumes which is large enough to host the cosmoDC2 lightcone. To prevent duplicate structures from being projected atop of one another, we follow the approach of previous sky simulations \citep{Blaizot:2003av, Kitzbichler:2006ec, bernyk2016}, and choose to randomly rotate each box replication.

While this strategy does decorrelate particle pairs at each box edge, we only need to replicate the Outer Rim volume  once per axis at a depth of $z\approx1.3$, so the impact of this decorrelation is relatively minimal. Though, in principle, this effect is present in the results shown in Figure~\ref{fig:lc_pk}, we find it to be negligible in practice. We refer the reader to \citet{Blaizot:2003av} for an in-depth study of these and other considerations related to generating lightcones from cosmological simulations.

\subsubsection{Halo Lightcone}
\label{sec:hlc}

To construct a catalog of halos on the observer's lightcone, an intuitive solution would be to re-run the halo-finding algorithm on the Outer Rim particle lightcone. Doing so, however, would be excessive in terms of computational cost, given that we already have a pre-constructed snapshot-based FOF halo catalog and associated merger tree (Sec.~\ref{sec:halos}). Therefore, we choose to pass the contents of the halo merger tree through the lightcone solver as described in Sec.~\ref{sec:plc}, resulting in a halo data set which is spatially commensurate with the particle lightcone.

In implementing the lightcone solver for the halo case, we use the same simulation volume replication and rotation strategy as previously described, and we also adopt the same linear approximation when interpolating spatial positions between snapshots. However, when generating halo lightcones, we conduct the interpolation proceeding backwards in time, interpolating halo positions in the direction of increasing redshift.

To understand why this time reversal is performed, it is helpful to imagine a branch of the halo merger tree spanning some time interval, which the lightcone surface ``slices through'' near the time of a halo merger event (shown in Figure~\ref{fig:halo-interp}; see Appendix~\ref{sec:lc_appdx} for more detail). We see that the adjacent snapshots $j$ and $j+1$ (grey planes), which bound the intersection of the lightcone and the merger tree branch, each host different objects -- the extent of our knowledge is that a merger happened \textit{somewhere} in the interval $t_{j-1} < t_\mathrm{merge} < t_{j}$.

Various prescriptions for assigning $t_\mathrm{merge}$, and defining halo properties at that time, have been described in the literature. For example, in building their Millennium-XXL-based lightcone, \citet{Smith:2017tzz} choose the merger time randomly per halo progenitor, and interpolate masses between snapshots. For cosmoDC2, we take a simpler approach, and assume that the merger has \textit{always} happened prior to it intersecting the lightcone surface (that is, for a merging merger tree branch that crosses the lightcone at $t_e$, we assert $t_\mathrm{merge} < t_e$ in all cases). We set the position of each halo within the lightcone by interpolating between the current halo position and that of its most massive progenitor, retaining all halo properties (mass, radius, etc.) as they appear in the later snapshot at $t_{j+1}$.

\begin{figure}[!ht]
  \centering
  \includegraphics[width=\linewidth]{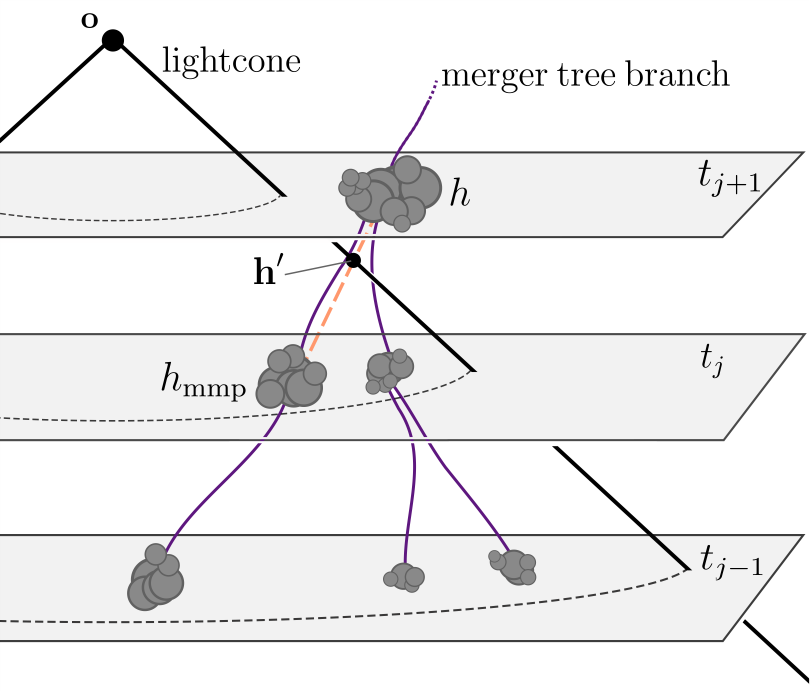}
\caption{Schematic of the interpolation process which fills the cosmoDC2 halo lightcone. Each plane represents a projected simulation snapshot, and time increases vertically, with the observer located at \textbf{o}. A merger tree branch including halo $h$ is seen crossing the observer's lightcone between snapshots $j$ and $j+1$ (the purple worldlines of each halo are unknown between the snapshots). Interpolation between halo $h$ and its most massive progenitor $h_\mathrm{mmp}$ (orange dashed line) is used to solve for the temporal and spatial components of event $\textbf{h}'$, where we place an object with properties (mass, etc.) identical to halo $h$.}
\label{fig:halo-interp}
\end{figure}

\subsection{Workflow}

\begin{figure}
\centering
\includegraphics[width=8cm]{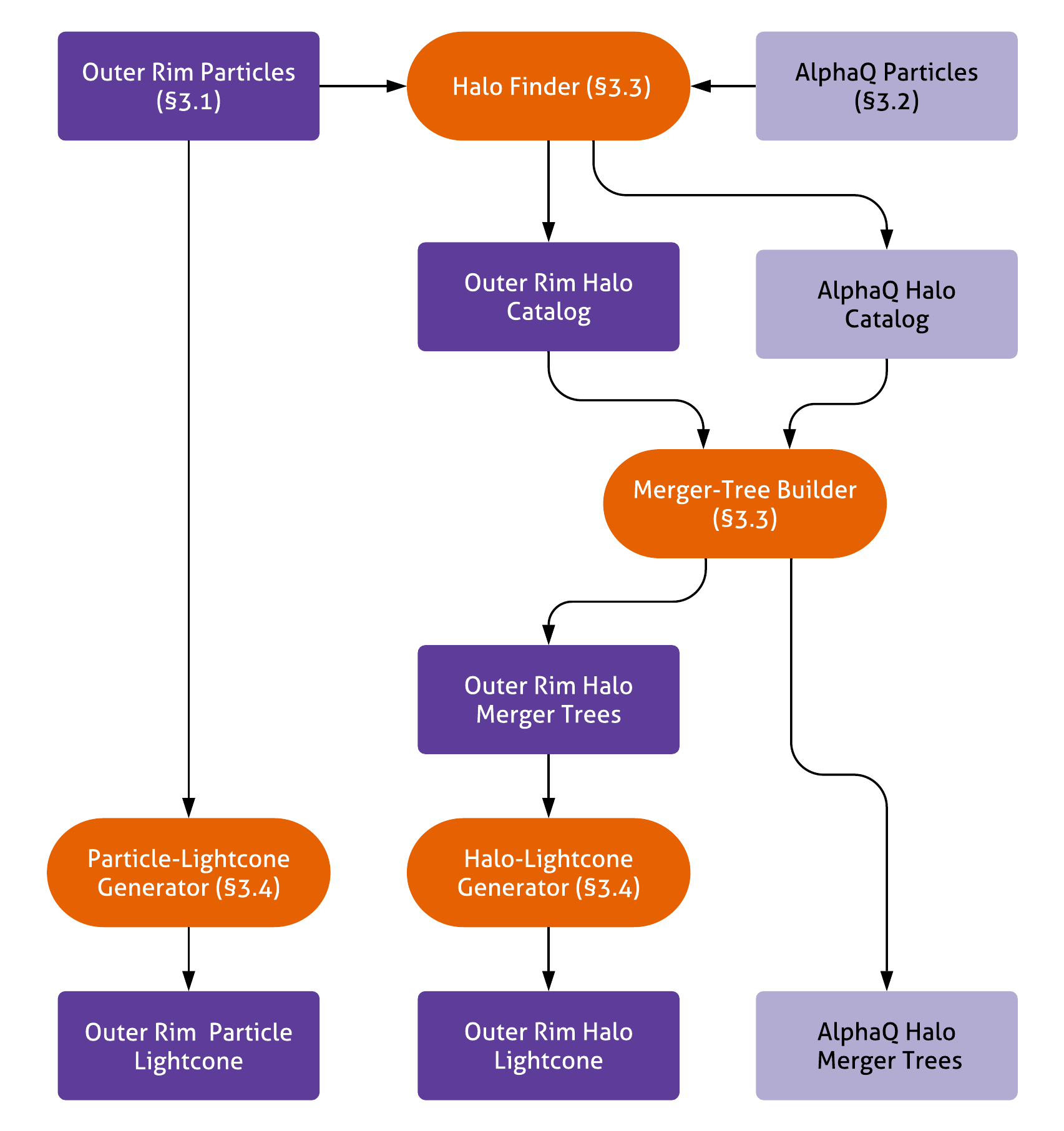}
\caption{Workflow to produce the Outer Rim and AlphaQ simulation data products used as inputs to the cosmoDC2 production pipeline. Data products are shown as rectangles in dark and light purple for data derived from the Outer Rim and AlphaQ simulations, respectively. Code modules are shown as ovals in dark orange. Numbers in parentheses refer to the sections in the paper where a detailed description of the workflow component is given.
}
\label{simulation_workflow}
\label{fig:simulation_workflow}
\end{figure}

Having described the simulations and the data products that are generated, we now provide a final summary by discussing the workflow for producing the inputs to the cosmoDC2 production pipeline. The workflow diagram is shown in Figure~\ref{simulation_workflow} and begins with the particle catalogs from the smaller AlphaQ  simulation
and the larger Outer Rim simulation.
These are both processed by the halo finder to construct halo catalogs which are then input into the merger tree builder. In the case of the Outer Rim simulation, the merger trees are used to build halo lightcones (see Sec.~\ref{sec:lightcone}) that serve as inputs for the Empirical-Model Pipeline and provide host halos for the galaxies in cosmoDC2. For the AlphaQ simulation the merger trees are used as inputs to the Galacticus SAM that is subsequently used to build the Galacticus Library.  The particle snapshots from the Outer Rim simulation are also input into the particle-lightcone generator to produce the inputs required for the Lensing Pipeline.

\section{Weak Lensing}
\label{sec:shear}

Weak lensing distortions are key observables of the LSST survey, providing constraints on the growth of cosmic structure and therefore dark energy \citep[e.g.,][]{Mandelbaum2018}. These distortions, which take the form of an isotropic change in area  (convergence) and an area-preserving change in shape (shear), can be mimicked in simulations by following the paths of photon rays as they traverse the matter field. In practice, maps of the lensing quantities are obtained as follows: the particle lightcone is divided into discrete shells, then photon paths are traced backwards in time from an observer grid to a `source' shell, with deflections applied corresponding to the surface density of particles at each `lens' shell between the source and observer using a ray-tracing algorithm \citep[e.g.,][]{2008ApJ...682....1D, 2009A&A...499...31H}.

\begin{figure}
\centering

\includegraphics[width=3.5cm]{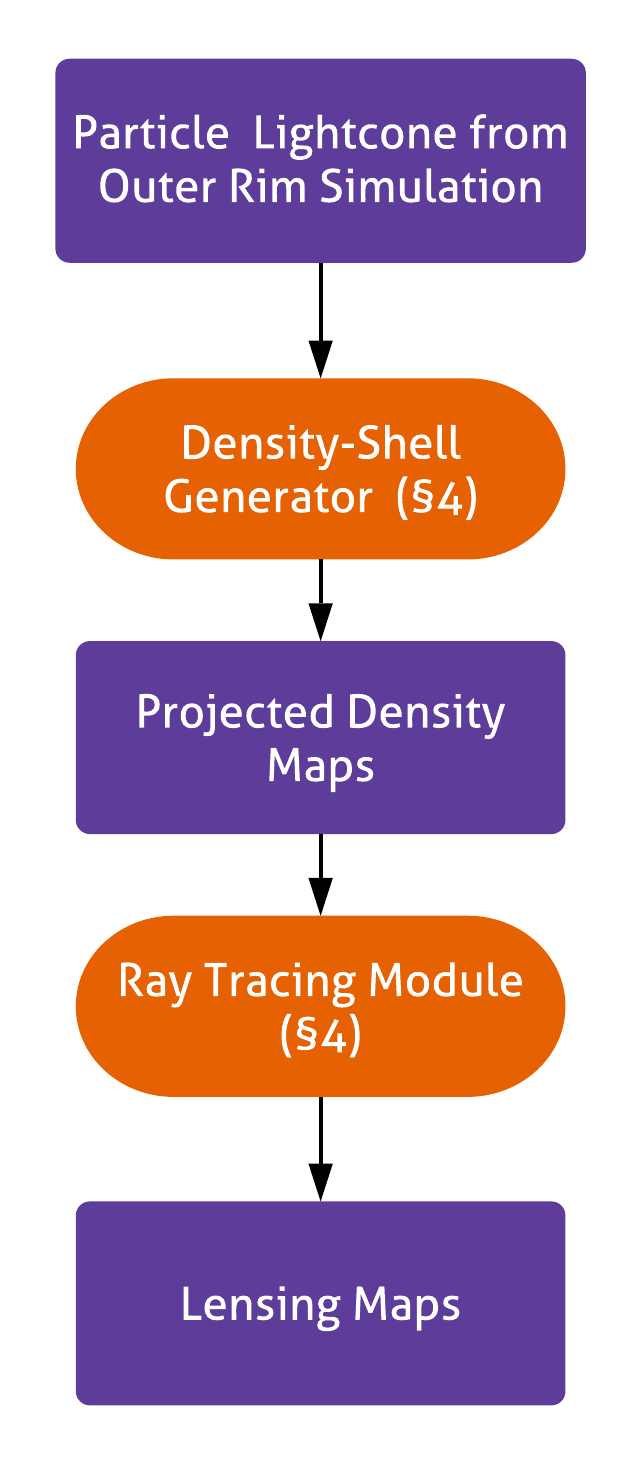}
\caption{Workflow for the shear pipeline.
Data products derived from the Outer Rim simulation are shown as boxes in dark purple and code modules are shown as ovals in dark orange.
}
\label{fig:shear-pipeline}
\end{figure}

\begin{figure}
  \centering
  \includegraphics[width=8.5cm]{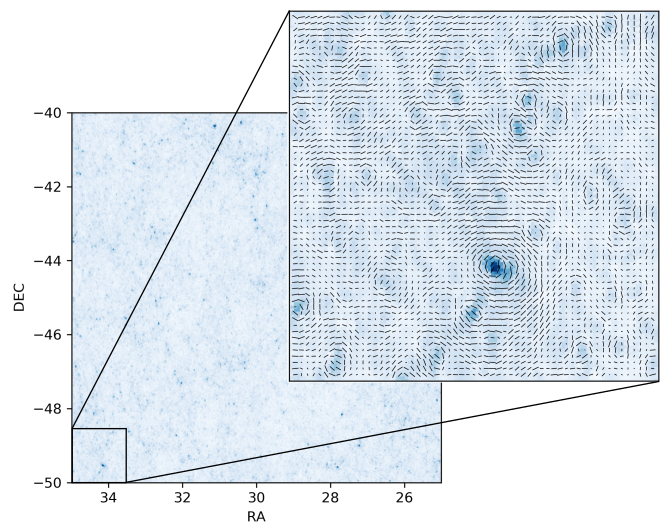}
\caption{Left image: Cartesian projection of a patch of the convergence source plane, in observer coordinates, at $z \approx 1.0$. Right image: Zoom-in of a box within this patch, with the cosmic shear field overlaid. For visualization purposes the lengths of the shear vectors are truncated to a maximum value above $\vert \gamma \vert = 0.025$.}
\label{fig:lensing_map}
\end{figure}

\begin{figure}
\centering
\includegraphics[width=\linewidth]{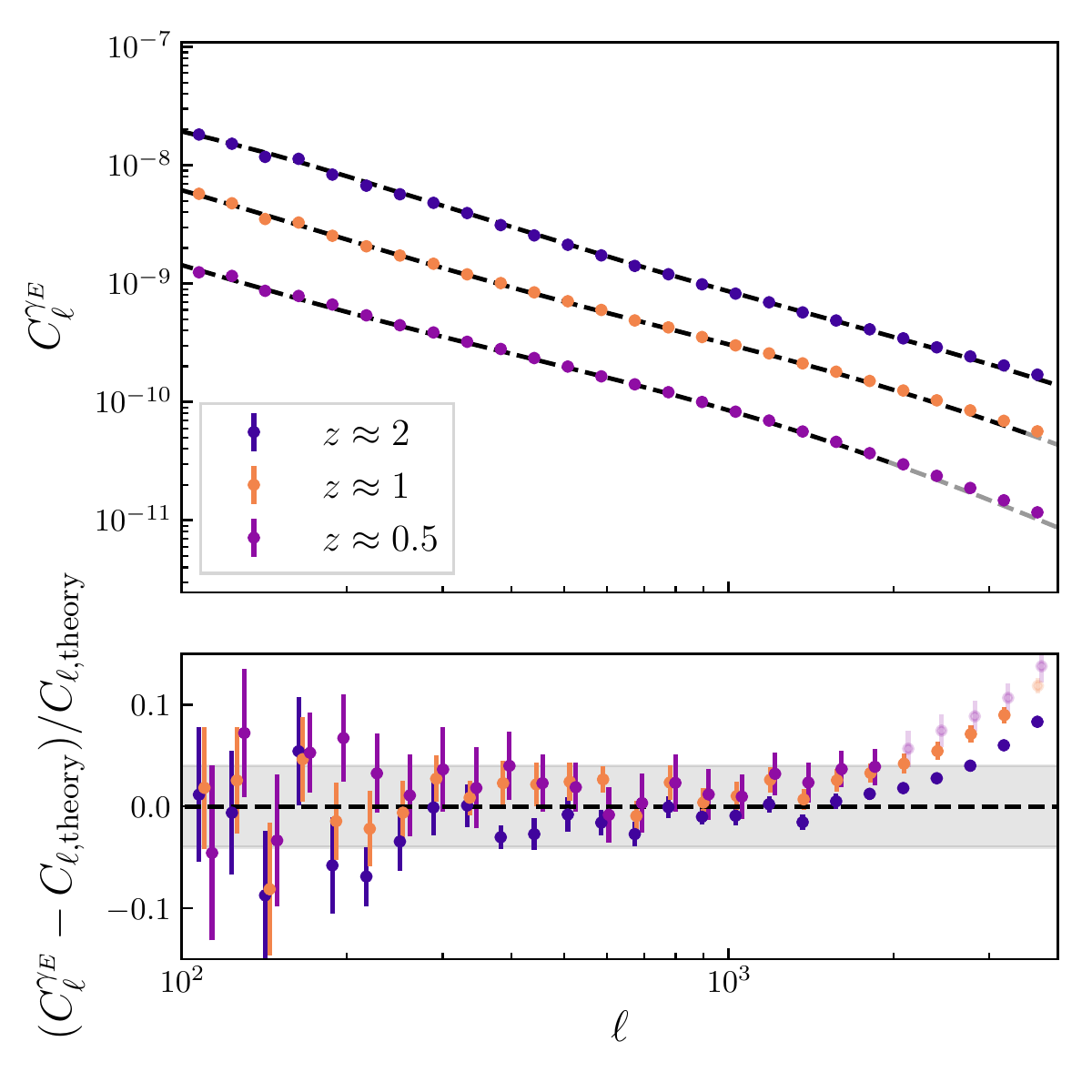}
\caption{E-mode shear power spectrum (top) and residuals with respect to theory (bottom) for a selection of source planes. Theory curves (black dashed lines) are computed using the Born approximation on the CosmicEmu power spectrum emulator, extrapolated to high $k$ values using Pad\'e approximants; the residual points are slightly offset for visualization purposes. The gray shaded region in the residual panel indicates the advertised $4\%$ accuracy of the power spectrum; transparency of the theory curves and residual points indicates that high-$k$ extrapolation accounts for more than $10\%$ of the total power in the theory curve.
}
\label{fig:shear-ps}
\end{figure}

 The baseDC2 lensing maps are built with the pipeline presented in P. Larsen et al. (2019, in preparation). The full workflow is illustrated in Figure~\ref{fig:shear-pipeline}. After we create a downsampled particle lightcone using the techniques described in Sec.~\ref{sec:plc} and divide it into discrete shells, we compute the surface densities on a HEALPix\footnote{\url{https://sourceforge.net/projects/healpix/}}~\citep{healpix} grid of Nside=4096 using a modified Delaunay Tesselation Field Estimator based on the code of \citet{rangeldtfe}. We then conduct ray tracing using the standard iterative equations of \cite{2009A&A...499...31H},
computing gradients and applying deflections on the full-sky maps with Lenspix routines \citep{lenspix}.\footnote{In particular we compute derivatives of spherical harmonics  using the {\tt HealpixAlm2GradientMap} routine, and deflect the mass shell to the observed plane using the {\tt HealpixInterpLensedMap} routine; these functions use cubic interpolation after a high resolution equi-cylindrical grid repixelization, with a cutoff in multipole of $\ell_{\rm{max}} = 8000$.} The shells are chosen to cover the line-of-sight distance between adjacent simulation outputs, as described in Sec.~\ref{sec:sim_or}, with a median width of approximately $114 \rm{Mpc}$.
Figure~\ref{fig:lensing_map} displays a cartesian projection of a $100$  $\rm{deg}^2$ patch of a resulting convergence source map at $z\approx 1$. The inset shows a sub-region with an expanded scale and with the shear field overlaid to show the tangential shearing around massive structures.  Figure~\ref{fig:shear-ps} compares the cosmic shear E-mode power spectra in three source maps to theoretical expectations; power spectra in the source maps are computed using the Polspice code with error bars obtained from jackknife sampling; theory curves are derived from the CosmicEmu power spectrum emulator using the Born approximation. These appear to agree to within the $4\%$ model errors anticipated from the power spectra on scales below $\ell \approx 2000$, and to within $10\%$ below $\ell \approx 4000$. As described in Sec.~\ref{sec:plc}, we note that cosmic shear on small scales is affected by interpolation of the downsampled particle lightcones, as well as limitations of the theoretical model and density estimation, and so the $\lesssim10\%$ level of agreement shown in Figure~\ref{fig:shear-ps} is expected.

Galaxies are assigned the lensing quantities of their source shell. We note that the maps resulting from the iterative ray-tracing equations of \citet{2009A&A...499...31H} are on an observer grid, so that galaxies must be shifted to their observed positions via the lensing deflections before computing the distortions. As the total deflection angles are small compared to the scale of pixelization, we use a first-order approximation to shift the positions from the source shell to the observed grid, given by  $\hat{n}_{\rm{obs}} \approx \hat{n}_{\rm{source}} - \nabla \phi (\hat{n}_{\rm{source}})$. We then assign lensing quantities by bilinear interpolation to the source maps.

Several complications arise when making simulation-based lensing predictions with ray-tracing techniques. These include the effects of shot noise, noise associated with triangulation in the density estimation of the downsampled particle data, particle lightcone interpolation errors, ringing effects from the use of spherical harmonics, and artificial smoothing due to pixelization and finite shell-width effects. We give a detailed discussion of these and other issues in  P. Larsen et al. (2019, in preparation).

\section{The Galaxy Catalog}
\label{sec:gc}
In this section we describe the method by which we populate the halo lightcone of the Outer Rim simulation with galaxies. First we produce the baseDC2 catalog by resampling galaxies from the UniverseMachine catalog and applying additional empirical modeling. Next we incorporate the weak lensing properties determined from the Outer Rim particle lightcone. Finally, we match the empirically-modeled galaxies to those in the Galacticus library to complete the assignment of the complex set of properties.
\subsection{Empirical Galaxy Catalog Generation}
\label{subsec:empirical}

Here we describe the ingredients of the empirical model that we use to construct the baseDC2 catalog. The  model parameters are simulation dependent and do not have a simple connection to observables. They have been chosen by an iterative procedure to produce a galaxy catalog whose distributions of observable properties are sufficiently close to that of the observational data to pass the DC2 validation criteria. It is a non-trivial task to calibrate these model parameters because they describe the behavior of rest-frame quantities in the baseDC2 model, rather than the observer-frame quantities used in the validation tests. For example, the validation test for galaxy number density as a function of magnitude (see Sec.~\ref{subsec:dndmag}) impacts the parameters used for modeling ultra-faint galaxies to be discussed in Sec.~\ref{subsec:faint}.  All of the model-parameter values can be obtained from the public code release.\footnote{\url{https://github.com/LSSTDESC/cosmodc2}}

\subsubsection{Restframe Colors}
\label{subsubsec:restframe_colors}

The starting point for baseDC2 is the publicly available UniverseMachine synthetic galaxy catalog \citep{behroozi_etal18}. The UniverseMachine is an empirical model for predicting the star-formation history of galaxies; the model is predicated upon the assumption that the mass assembly of a galaxy is correlated with the assembly of its underlying dark matter halo. While this is a longstanding assumption of the semi-analytic modeling approach to galaxy formation \citep[e.g.,][]{1993MNRAS.264..201K,1999MNRAS.310.1087S,benson_2010b,somerville_dave15}, recent theoretical developments have enabled a new generation of models to leverage this assumption in a way that significantly improves the complexity that can be captured with empirical techniques \citep{becker15,cohn17,moster_etal17,rodriguez_puebla_etal17}. The  UniverseMachine model has been shown to faithfully capture a wide range of statistics summarizing the observed galaxy distribution across redshift, including stellar mass functions, quenched fractions, and the SFR-dependence of two-point clustering.

The UniverseMachine catalog we use contains synthetic galaxies populating snapshots in the MDPL2 simulation \citep{klypin_etal16}, such that every subhalo identified by Rockstar~\citep{2013ApJ...762..109B} in MDPL2 is populated with a synthetic galaxy. For the purposes of baseDC2, we restrict attention to just two attributes of these synthetic galaxies, stellar mass $\mstar,$ and star-formation rate $\sfr.$ In particular, the value of $\mstar$ that we use is defined as the total surviving stellar mass belonging to the galaxy, excluding contributions from intra-cluster light.  We use the GalSampler technique to transfer the UniverseMachine galaxy population in MDPL2 to the Outer Rim simulation (for technical details, see A. Hearin et al, 2019 in preparation). Briefly, for every host halo in the Outer Rim, we randomly select a host halo in MDPL2 of similar mass, and map the galaxy content of the selected MDPL2 halo into the Outer Rim halo, preserving the halo-centric positions and velocities of the galaxies. By construction, the GalSampler technique preserves the conditional distribution $P(\sfr, \mstar\vert\mhalo),$ as well as the halo mass dependence of the UniverseMachine halo occupation statistics, $P(N_{\rm gal}\vert\mhalo).$  For the most massive halos in the Outer Rim simulation, which have no counterparts in MDPL2, using the GalSampler random selection procedure ensures that we are not repeatedly resampling galaxies from the same MDPL2 halo. We take the larger halo mass of the Outer Rim halo into account by applying a redshift-dependent boost to the UniverseMachine value of $\mstar$.

At this stage, every halo in the Outer Rim lightcone has been populated with synthetic galaxies with $\mstar$ and $\sfr.$ We model restframe absolute magnitude $\magr$ as a function of $x\equiv\log_{10}M_{\star}$ and redshift $z,$ we map $M_r$ onto every synthetic galaxy using the following model:

\begin{equation}
\label{eq:magr}
\langle\magr\vert\mstar,z\rangle = \left[M_{r}^{0} - x\alpha(x)\right]\times\left(1 + f(x, z)\right),
\end{equation}
where $M_{r}^{0}=-20.1$ is a constant, and where we model both $\alpha(x)$ and $f(x, z)$ using $S(x),$ a sigmoid function:
\begin{equation}
\label{eq:sigmoid}
S(x) = y_{\rm min} + \frac{y_{\rm max}-y_{\rm min}}{1 + \exp(-k(x-x_0))}.
\end{equation}
The function $\alpha(x)$ controls the $\mstar$-dependence of the power-law slope. For $\alpha(x),$ we use a low-mass slope of $y_{\rm min}=1.75$ and a high-mass slope of $y_{\rm min}=1.8,$ with a transition speed of $k=2.5$ and a pivot mass of $\log_{10}M_{\rm pivot}/\rm{M}_{\odot}\equiv x_0=10.$

The function $f(x, z)$ controls the redshift evolution of the $\langle\magr\vert\mstar,z\rangle$ relation. Because galaxies at higher redshift are generally composed of younger stellar populations, we expect that the median value $\langle\magr\vert\mstar,z\rangle$ brightens with redshift, and that this brightening is stronger for lower-mass galaxies. We capture this complexity by modeling
\begin{equation}
f(x, z) = \frac{\delta(x)}{1 + \exp(-k(z-z_0))},
\end{equation}
with a transition speed $k=10,$ and a pivot redshift $z_{0}=0.7,$ and a third-order polynomial for $\delta(x)$ defined to pass through the points $\{(6, -2), (8, -1.5), (10, -0.5), (12, 0)\}.$ The preceding pairs of numbers give values for  $(\log(\mstar),  \langle\Delta M_r\rangle)$, where $\langle\Delta M_r\rangle$ is the average brightening for galaxies for a given stellar mass.

Having mapped $\mstar$ and $\magr$ onto every synthetic galaxy, we proceed to model restframe colors $g-r$ and $r-i$. For each distribution we use a double Gaussian, with statistically distinct star-forming and red sequence populations:

\begin{eqnarray}
\label{eq:double_gaussian}
P({g-r}\vert\magr, z) &=& F_{\rm q}(\magr, z)\times\mathcal{N}^{\rm q}_{\rm g-r}(\mu; \sigma) \nonumber \\
 &+&  \left(1-F_{\rm q}(\magr, z)\right)\times\mathcal{N}^{\rm sf}_{\rm g-r}(\mu; \sigma),
\end{eqnarray}
and likewise for $P(r-i\vert\magr, z).$ We model the two-dimensional dependence of $F_{\rm q}(\magr, z)$ using a composition of sigmoid functions:
\begin{equation}
F_{\rm q}(\magr, z) = F^{z_{0}}_{\rm q}(\magr) + \frac{F^{z_1}_{\rm q}(\magr)-F^{z_0}_{\rm q}(\magr)}{1 + \exp(-k_{\rm z}(z-z_{\rm c}))},
\end{equation}
where for both $g-r$ and $r-i$ colors we use $z_{\rm c}=0.5$ and $k_{\rm z}=12.$ The $F^{z_{\rm i}}_{\rm q}$ functions characterize the $\magr$-dependence of the quenched fraction at the two asymptotic redshifts, $z_0$ and $z_1.$ We model $F^{z_{\rm i}}_{\rm q}(\magr)$ as
\begin{eqnarray}
F^{z_{\rm i}}_{\rm q}(\magr) = f_{\rm q}^{\rm faint} + \frac{f_{\rm q}^{\rm bright}-f_{\rm q}^{\rm faint}}{1 + \exp(-k_{\rm r}(\magr-\magr^{\rm c}))}.
\end{eqnarray}

In cosmoDC2, the widths of the quenched and star-forming sequences are constant, but the centroids depend on both mass and redshift, i.e., in Eq.~\ref{eq:double_gaussian}, for each sequence and broadband color, $\mu=\mu(\magr, z).$ We model this simultaneous dependence as a composition of sigmoid functions.
\begin{equation}
\label{eq:mu}
\mu(\magr, z) = \mu_{z_{0}}(\magr) + \frac{\mu_{z_{1}}(\magr)-\mu_{z_{0}}(\magr)}{1 + \exp(-\alpha_{\rm z}(z-z_{\rm c}))},
\end{equation}
where
\begin{equation}
\mu_{z_{i}}(\magr) = \mu_{\rm faint} + \frac{\mu_{\rm bright}-\mu_{\rm faint}}{1 + \exp(-\alpha_{\rm r}(\magr-\magr^{\rm c}))}.
\end{equation}

We arrived at these functional forms and best-fit values after considerable experimentation and iteration with the DESCQA color validation tests to be presented in Sec.~\ref{subsec:color}.

\subsubsection{Cluster Environment}

While the methods used in Sec.~\ref{subsubsec:restframe_colors} produce model galaxies with realistic stellar mass and broadband flux, as well as reasonably accurate two-point clustering, two additional ingredients are needed in order to meet validation requirements in cluster environments. First, more stellar mass is required of central galaxies in very massive halos, $\mhalo\gtrsim10^{14}\msun$\citep{huang_etal18}. Second, the normalization of the mass-richness relation in cluster-mass halos appears to be $\sim20\%$ low relative to expectations based on DES data. In principle, the UniverseMachine model could capture these effects if suitable observational data for galaxies in cluster environments were used to tune the UniverseMachine model parameters. In practice, the necessity of these modifications is not surprising because the observational constraints used to fit the UniverseMachine model are relatively insensitive to the behavior of the galaxy-halo connection in the statistically rare environment of very massive halos.

To address the boost to the stellar mass of cluster centrals, we remapped the $\langle\mstar\vert\mhalo\rangle$ relation for $\mhalo>\mhalo^{c}=10^{13.5}\msun$ according to $\mstar\propto\left(\mhalo/\mhalo^{c}\right)^{\alpha},$ using $\alpha=0.65,$ and set the normalization according to the existing value for $\langle\mstar\vert\mhalo^{c}\rangle.$ To boost the mass-richness relation, we generate an additional Monte Carlo realization of cluster satellites so that the total number of objects increases by $20\%$ in all halos $\mhalo>\mhalo^{c},$ decreasing this boost factor linearly in $\log\mhalo$ so that the boost is zero for halos with $\mhalo\leq10^{13}\msun.$

\subsubsection{Ultra-Faint Galaxies}
\label{subsec:faint}

A range of science goals related to weak lensing and deblending benefit from a synthetic catalog that is complete down to galaxy masses below the resolution limit of the simulation. As discussed in Sec.~\ref{subsec:dndmag}, the primary DESCQA validation requirement that drives the need for this additional modeling is quantified in the test for the cumulative galaxy number counts as a function of apparent magnitude. To meet this requirement, part of the cosmoDC2 model includes a population of ``ultra-faint galaxies" that are disconnected from resolved halos in the Outer Rim simulation.

Our modeling for the ultra-faint population begins by defining how many galaxies should be included in order to boost the abundance of faint galaxies to meet the DESCQA validation criteria. 
We make two physical assumptions to determine the abundance of ultra-faint galaxies at each redshift. The assumptions are that there is one-to-one correspondence between galaxies and (sub)halos, accounting for both distinct host halos and for subhalos within them,
and that the (sub)halo mass function $\dndmpeak$ should exhibit power-law behavior at the low-mass end.
These assumptions are used to extend the reach of the simulation in the following way. Let $\mpeak^{\rm lim}$ denote the resolution limit of the simulation, i.e., below $\mpeaktwo{lim}$ we begin to see departures from power-law behavior of $\dndmpeak.$ Then for any particular value $\mpeaktwo{ext}<\mpeaktwo{lim},$ we can simply extrapolate the power-law approximation to $\dndmpeak$ to estimate how many (sub)halos are missing due to the finite mass resolution of the simulation. We elaborate upon this procedure below.

In each snapshot of the MDPL2 subhalo catalogs used in baseDC2, we identify a value of $\mpeak$ that is sufficiently larger than $\mpeaktwo{lim},$ and fit a power law to $\dndmpeak$ in the neighborhood of this mass. Selecting $\mpeaktwo{ext}=10^{9.8}\rm{M}_{\odot},$ we then calculate $n_{\rm ultra-faints}(z\vert\mpeak>\mpeaktwo{ext}),$ the cumulative comoving number density of (sub)halos that would be present if MDPl2 were $\mpeak$-complete to this mass:
\begin{equation}\label{eq:ultrafaintdensity}
n_{\rm ultra-faints}(z\vert\mpeak>\mpeaktwo{ext}) \equiv \int_{\mpeaktwo{ext}}^{\mpeaktwo{lim}}{\rm d}\mpeak\dndmpeakfit(z)
\end{equation}
Equation \ref{eq:ultrafaintdensity} defines the volume number density of synthetic ultra-faint galaxies to which we will apply the GalSampler technique to add them to baseDC2. We also use $\dndmpeakfit$ to draw values of $\mpeak$ for each ultra-faint galaxy we add. Once we have a sample of synthetic ultra-faint galaxies that have values of $\mpeak,$ we assign stellar masses according to a power-law fit to the faint end of the UniverseMachine relation $\langle\mstar\vert\mpeak, z \rangle,$ and we assign uniform random values to $P(<{\rm SFR}\vert\mstar)$, the CDF of the \sfr~conditional probability distribution. Based on the assignments of $\mstar$ and $\sfr$, we assign colors as described in Sec.~\ref{subsubsec:restframe_colors}. We then assign spatially random locations to these galaxies, computing redshifts from the corresponding comoving distances. At this point, the synthetic ultra-faint galaxies have all the attributes needed to treat them as ordinary UniverseMachine galaxies in the baseDC2 pipeline.

As in the case of the rest-frame color model, the parameter values presented here are obtained by boosting the number of ultra-faint galaxies until there are sufficiently many to pass the cumulative number-density validation test to be presented in Sec.~\ref{subsec:dndmag}. In the future, for use-cases requiring more realistic spatial distributions of the ultra-faint population, it would be necessary to incorporate expected correlations between the density field and the positions of very low-mass galaxies.

\subsubsection{HOD Comparison with UniverseMachine}
\label{subsec:hod}

One of the simplest ways to quantify the relationship between galaxies and the cosmic density field is through the Halo Occupation Distribution (HOD), $P(N_{\rm gal}\vert M_{\rm halo}),$ the probability that a halo of mass $M_{\rm halo}$ hosts $N_{\rm gal}$ galaxies that meet some selection criteria. In order to demonstrate that the $P(N_{\rm gal}\vert M_{\rm halo})$ in cosmoDC2 is reasonably realistic, we compare our HOD to that seen in mock catalogs made with the UniverseMachine. Two-point projected clustering in the UniverseMachine model has been shown to exhibit close agreement with SDSS \citep{behroozi_etal18}, and so for purposes of ensuring reasonably accurate correlations between galaxies and the density field, we compare our redshift-zero HOD to that seen in UniverseMachine.

\begin{figure}
\centering
\includegraphics[width=8cm]{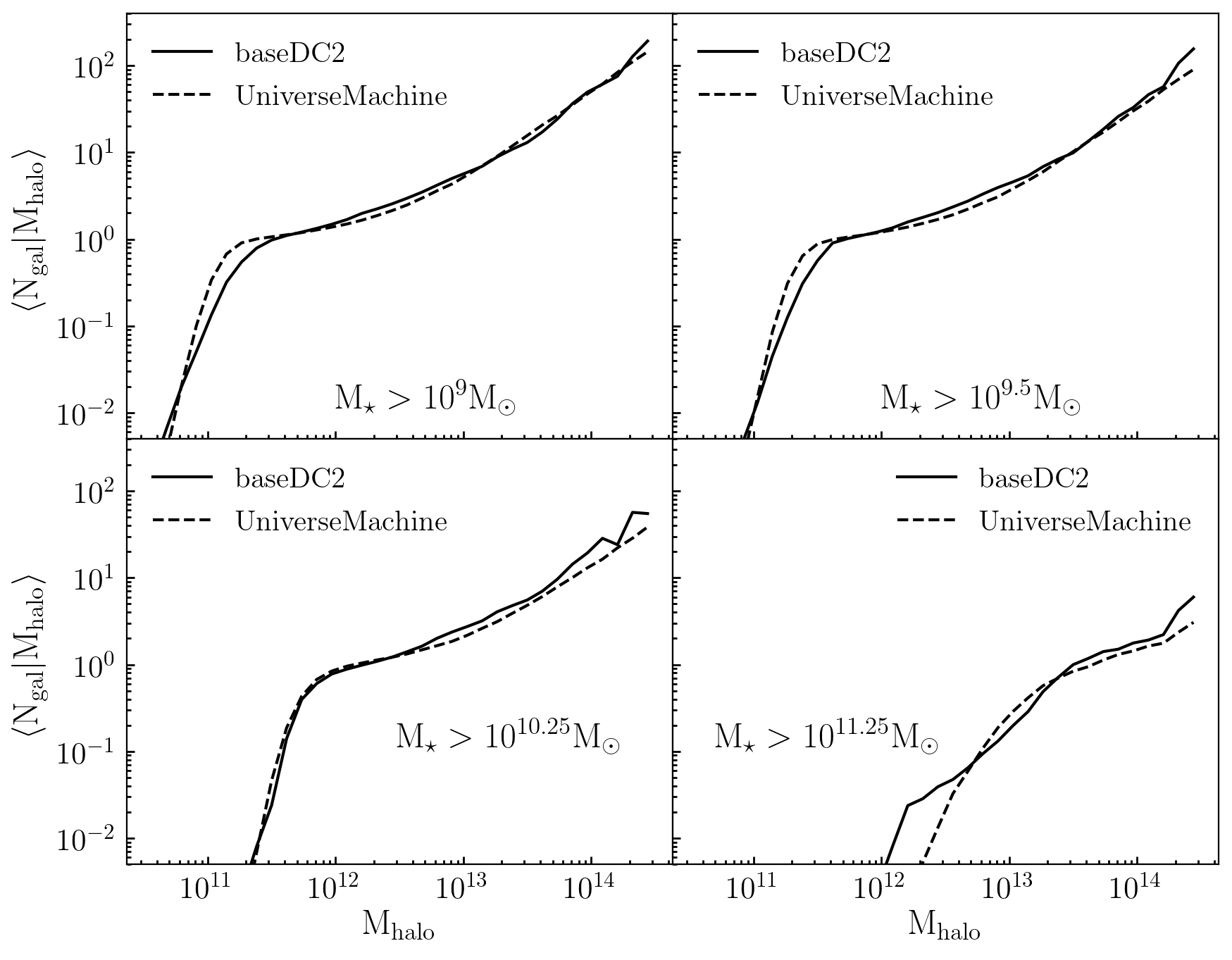}
\caption{Comparison of the $z=0.15$ HOD in the UniverseMachine mock (dashed curves) with those of baseDC2 (solid curves). The HOD quantifies $P(N_{\rm gal}\vert M_{\rm halo}),$ the probability that a halo of mass $M_{\rm halo}$ hosts $N_{\rm gal}$ galaxies with stellar mass $M_{\star}$ greater than the threshold indicated in the panel. The first moment of the HOD, $\langle N_{\rm gal}\vert M_{\rm halo}\rangle,$ is shown on the vertical axis of each panel as a function of $M_{\rm halo}.$ Solid curves show the corresponding HOD in the baseDC2 model that is the foundation of cosmoDC2. We tune our HOD to match UniverseMachine so that our model can inherit the observational realism of the galaxy-halo connection shown in \cite{behroozi_etal18}.
}
\label{fig:umhod}
\end{figure}

In Figure \ref{fig:umhod}, the dashed curves shows the HOD of $z=0.15$ UniverseMachine galaxies, with different stellar mass thresholds as indicated in each panel. To calculate the corresponding quantity in our model, we use the $z=0.15$ snapshot of the Outer Rim halo catalog populated with baseDC2 galaxies. The good agreement between the dashed and solid curves in each panel of Figure \ref{fig:umhod} should be sufficient to ensure that cosmoDC2 has reasonably accurate relationships between stellar mass and halo mass. Since stellar mass and luminosity are tightly correlated in the cosmoDC2 model (see Sec.~\ref{subsubsec:restframe_colors}), then the HOD in cosmoDC2 will naturally inherit dependence upon broadband magnitude. As discussed in Sec.~\ref{sec:results}, our tuning of this technique was fairly coarse relative to the accuracy with which the HOD has been shown to recover the clustering of specific galaxy samples \citep[e.g.,][]{zheng07}. To build the cosmoDC2 model, we instead prioritized modeling galaxies with HODs that exhibit the expected scaling with a wide variety of complex observational selection functions, and we have found the level of agreement shown in Figure \ref{fig:umhod} to be sufficient to serve a broad range of science applications of DC2.

\subsection{The Galacticus Library}
\label{subsec:sam}

The Galacticus SAM solves, along each branch of an input merger tree, a set of ODEs which describe the evolution of coarse-grained properties of the galaxy that forms within that branch. This differential evolution is supplemented with impulsive events that describe galaxy mergers. In addition to merging, the physics modeled by Galacticus also includes the cooling of gas in the circum-galactic medium and its inflow into galaxies, star formation, supermassive black
hole growth, feedback processes powered by both supernovae and active galactic nuclei, and metal production. Galacticus follows the evolution of the disk and bulge components separately. The latter component forms as a result of galaxy mergers or via dynamical instabilities of the galactic disk. In either case, some fraction of the disk's mass is transferred to the bulge component.

	Our Galacticus library was generated by running version 0.9.4 of the Galacticus code\footnote{\url{https://bitbucket.org/galacticusdev/galacticus/wiki}} on merger trees built from the AlphaQ simulation.
	Galacticus outputs were requested at each snapshot redshift of the input merger trees so that a complete history of each galaxy would be available in the Galacticus Library.
	We use the default 0.9.4 parameter file to specify the inputs to the Galacticus model. Since these input parameters are tuned to observational data using extended Press-Schechter trees rather than merger trees from N-body simulations, we do not expect our results to match these observational data perfectly. However, as mentioned in Sec.\ref{sec:e-to-e}, the collection of Galacticus galaxies is serving as a library from which to select galaxies with suitable properties, so that a complete match to the data is not required.

	In addition to the basic properties such as stellar mass, star-formation rate, gas and stellar metallicity, the user can supply filters for which Galacticus then calculates luminosities. In principle, these luminosities can be determined by modeling the spectral-energy distribution (SED) of a galaxy with a stellar-population synthesis (SPS) model \citep{conroy-2009}, and integrating the resulting SED over the band-pass for the selected filter. The SPS model provides a library of single-stellar-populations (SSP) SEDs that depend on metallicity and time. The SED of a galaxy at a given redshift is obtained by convolving over time and metallicity, the \sfr~as calculated by Galacticus, with these SSP SEDs. In practice, it is too time consuming to calculate these integrals for each SED on the fly, so Galacticus pre-computes a table of integrals of the SSP SEDs over the desired filter band-passes and uses these as coefficients in the convolution integral with the SFR.
	Luminosities for the desired band passes are available in either rest or observer frame and are computed separately for the disk and bulge component of each galaxy. Observer-frame luminosities are calculated by appropriately blue-shifting the filter transmission function depending on the cosmological redshift of the galaxy. (Recall that output redshifts are specified so that these required blue-shifts are known in advance.) \ Note however, that the blue-shifts that account for the line-of-sight peculiar velocities are not included. Although the direct computation of observer-frame magnitudes obviates the need to perform any k-corrections on the rest-frame magnitudes, we still need to interpolate both sets of magnitudes to the galaxy lightcone redshifts. This will be discussed in Sec.~\ref{subsec:z-interp}.

	Galacticus luminosities are computed in the AB-magnitude system. We convert the rest-frame luminosities $L$ to magnitudes using $M= -2.5\log_{10}(L)$. Observer-frame luminosities are converted to apparent magnitudes by including additional factors of $-2.5\log_{10}(1+z)$ and $\mu(z)$, where $z$ is the redshift and $\mu$ is the distance modulus of the galaxy. These factors account for the compression of photon frequencies in the observer frame and the luminosity distance, respectively.

	The luminosities provided by these user-selected filters are critical for providing the galaxy properties required by the validation tests. For example, many of these tests make cuts on observer-frame LSST magnitudes, which are not obtained from empirical modeling.
	In addition to the LSST $ugrizy$ filters, we  included SDSS $ugriz$ filters to facilitate the validation of cosmoDC2 against SDSS data and the Johnson $B$ and $V$ filters to provide inputs for the image simulations.  We also define a set of 30 top-hat filters spanning the range from 100~nm to 2~$\mu$m. The width of these filters grow with wavelength but are designed to provide roughly constant resolutions  $\lambda/\Delta\lambda$ that vary from $\sim 4$ to $\sim 7$. These top-hat filters provide a coarsely binned estimate of the galaxy's SED that is based on the star-formation history of each galaxy. They are required inputs for the image simulations and are critical for evaluating the accuracy of photometric-redshift determinations. Finally, to estimate emission-line luminosities (described below), we add three continuum filters that compute the ionizing luminosity for \HI, \HeII, and \OII, respectively.

	Dust corrections and emission lines are added in post-processing. The dust model is from \cite{ferrara-1999} who used ray-tracing simulations to calculate dust-attenuation curves as a function of inclination, dust distribution and other dust properties. In post processing, a random inclination is generated for each galaxy and the attenuation for each filter is computed by interpolating (or extrapolating) the attenuation at the effective wavelength
	of the filter from the tabulated values.
	The effective wavelength is defined as
	$\int \lambda T(\lambda)d\lambda/ \int T(\lambda)d\lambda$,
	the wavelength averaged over the filter transmission, $T(\lambda)$.

	The emission-line model is based on \cite{panuzzo-2003}, and its specific implementation within Galacticus is described in detail by \cite{2018MNRAS.474..177M}. Using this model, emission line luminosities are computed using the photo-ionization code Cloudy \citep{cloudy} as a function of a grid of values for \HII~region densities, metallicities and the ionizing luminosities for \HI, \HeII, and \OII.  In post-processing, for each galaxy, the emission-line luminosities are interpolated from this pre-computed grid using the measured values of the ionizing luminosities that are supplied by Galacticus.

	As a final step, we remove galaxies with extreme values in the dust-corrected magnitudes, colors and total extinction from the Galacticus library. Any galaxies with any totally extincted dust-corrected magnitude (due to extrapolation failures in the dust model), values of the rest-frame $\rm{B}-\rm{V}$ excess color or the ratio of total to selective extinction, $R_V$,   close to zero, or values of  dust-corrected rest- or observer-frame colors or values of the V-band total extinction, $A_V$, falling outside of the limits shown in Table~\ref{tab:cuts} are cut from the library.

\begin{table}[!htb]
\begin{center}
\label{tab:cuts}
\caption{Color and $A_v$ cuts applied to the Galacticus library to remove unphysical galaxies}
\begin{tabular}{|c|c|c|}
\hline\hline
   Quantity  &  Lower Limit & Upper Limit\\ \hline
   Rest-frame $g-r$ & -0.5 &  1.5 \\
   Rest-frame $r-i$ & -0.5 &  1.5 \\
   Rest-frame $i-z$ & -0.5 &  2.0 \\
   Observer-frame $g-r$ & -0.5 &  2.5 \\
   Observer-frame $r-i$ & -0.5 &  2.0\\
   Observer-frame $i-z$ & -1.0 &  2.5 \\
   $A_V$ & -0.1 & 3.1 \\
   \hline\hline
\end{tabular}
\end{center}
\end{table}

\subsection{Galaxy Catalog and Galaxy Library Matching}
\label{subsubsec:library-match}

\begin{figure}
\centering
\includegraphics[width=8cm]{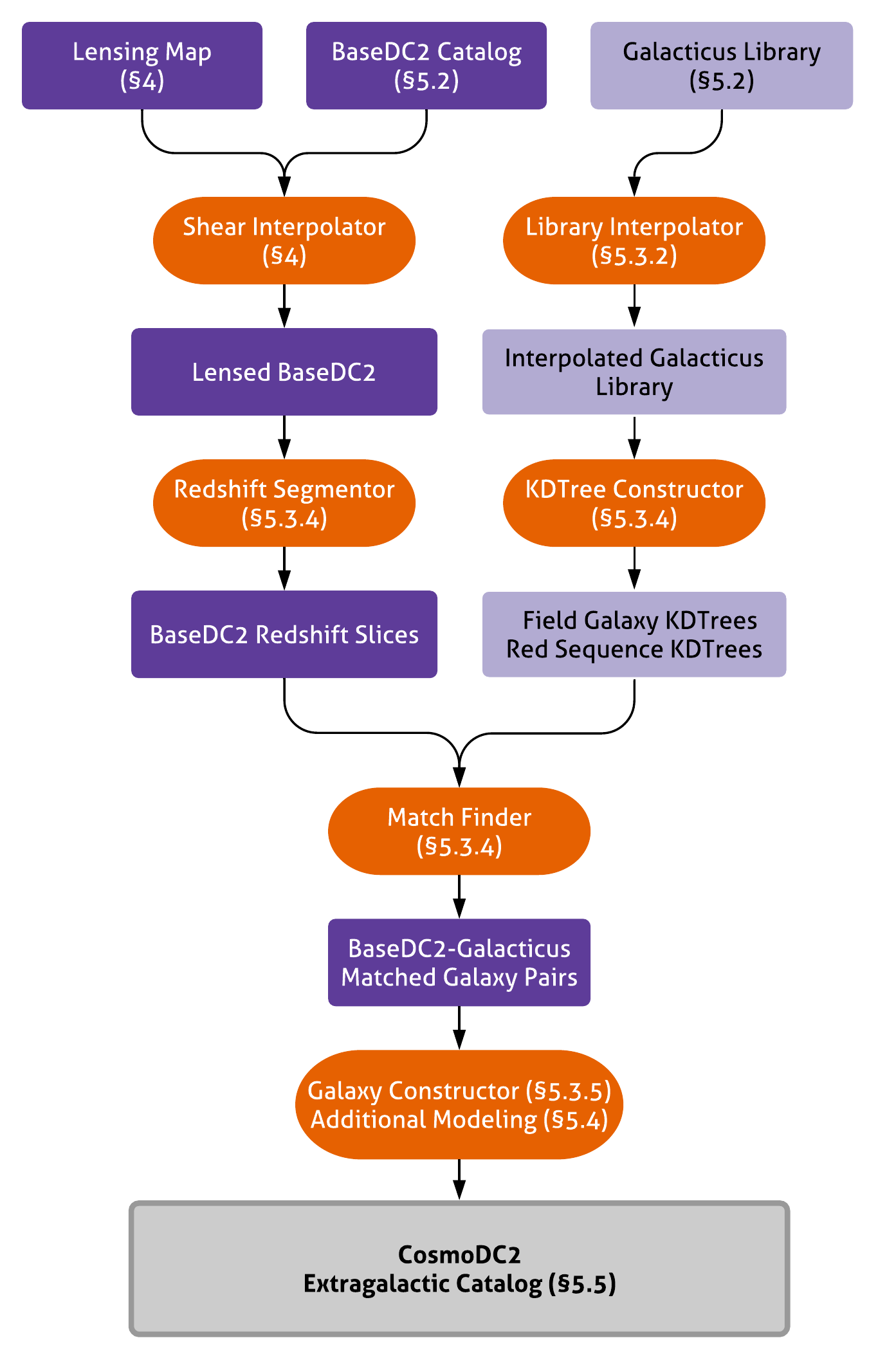}
\caption{ Workflow for the Galacticus match-up pipeline to produce the
  final output for cosmoDC2.  Data products are shown as rectangles in
  dark and light purple for data derived from the Outer Rim and
  auxiliary simulations, respectively. Code modules are shown as ovals
  in dark orange. }
\label{fig:matchup-pipeline}
\end{figure}

The goal of the Match-Up Pipeline, shown in Figure~\ref{fig:matchup-pipeline}, is to find a Galacticus library
galaxy for each baseDC2 galaxy that best preserves the careful tuning of
existing properties while incorporating additional information provided by the library.
The key galaxy properties to match between baseDC2 and the Galacticus library are rest frame $r$-band magnitude, and rest frame $g-r$ and $r-i$ color. All other baseDC2 properties, including stellar mass and SFR, are copied directly into cosmoDC2. Weak lensing properties (see Sec.~\ref{sec:shear}) are incorporated into the baseDC2 catalog before the matching.

\subsubsection{Quality of Match Metric}

The quality of the match between a library galaxy and a baseDC2 galaxy is measured by the Euclidean distance between their 3-dimensional property vectors $P^{\rm{lib}}$ and $P^{\rm{base}}$, respectively, where the components of $P$ are $M_r$, rest frame $g-r$ and $r-i$ color:
\begin{equation}
    D^2 = \sum_{j} (P^{\rm{lib}}_{j} - P^{\rm{base}}_{j})^2.
    \label{eq:matchup_dist}
\end{equation}

Cluster red sequence galaxies have a tight relationship between color
and redshift \citep{bower92,rozo_etal15c}.  Optical cluster finders such as redMaPPer \citep{rykoff_etal14b} rely on this tight relationship to help isolate
cluster members, and so in order to ensure that cosmoDC2 possesses a tight relation between redshift and observed galaxy color we include a second term in the distance calculation for cluster red sequence galaxies:
\begin{equation}
  D^2 = \sum_{j}^{} \left(P^{\rm{lib}}_j- P^{\rm{base}}_j\right)^2 + \sum_{j}^{} \left(Q^{\rm{lib}}_j - Q^{\rm{DES}}_j(z)\right)^2,
  \label{eq:matchup_dist_rs}
\end{equation}
where the color vector Q has components of ${g-r},\ {r-i},$ and ${i-z}$ in the observer frame, and $Q^{\rm{DES}}(z)$ is the expected mean red sequence color as a function
of redshift observed in DES \citep{rykoff_des_2016}. Cluster red sequence galaxies are defined as all red sequence galaxies with $M_{\rm halo}>10^{13}h^{-1}\rm{M}_{\odot}$. To ensure a smooth transition between the use of the two metrics, we also apply Eq.~\ref{eq:matchup_dist_rs} to a fraction of the galaxies for which $10^{12}h^{-1}\rm{M}_{\odot}<M_{\rm halo}<10^{13}h^{-1}\rm{M}_{\odot}$. This fraction increases log-linearly with halo mass from zero to one. We discuss this modeling choice further
in Sec.~\ref{sec:outlook}.

\subsubsection{Redshift Interpolation of Library Galaxies}
\label{subsec:z-interp}

Whereas baseDC2 is constructed on a lightcone, the Galacticus
library has been constructed at discrete redshift snapshots. If galaxies in the baseDC2 lightcone are naively matched to galaxies in the Galacticus library, then discrete
bands in observed-color and redshift space will be clearly visible. These
bands appear because the color distribution of galaxies changes noticeably between the redshifts of two adjacent snapshots. This is due both to redshifting of the galaxy SED and, to a lesser extent, evolution of the galaxy population. We use interpolation to compute the properties of galaxies lying between the snapshot redshifts. Our procedure substantially reduces the discreteness effects that would otherwise be present.

The Galacticus library contains the full history of each galaxy at every snapshot redshift. To shift a Galacticus galaxy to an intermediate redshift $z$, we linearly interpolate the properties of the galaxy between adjacent snapshots, expressed as:
\begin{equation}
  g(z) = 
    f(z_t) + \left(z-z_t\right)\frac{f(z_{t+1})-f(z_t)}{z_{t+1}-z_{t}},
  \label{eg:gal_prop_interpolation}
\end{equation}
where $z_t$ and $z_{t+1}$ are the discrete snapshot redshifts of the library that bracket $z$, $g(z)$ is the interpolated function of a galaxy
property and $f(z_{i})$ is the value of a galaxy library property at redshift $z_i$.

Not all galaxies in the library are suited for interpolation between snapshots. For example, some galaxies in Galacticus either formed recently or merged with another galaxy and are therefore missing from a snapshot; galaxies may have evolved in to or out of regions of
color-magnitude space that fail to pass quality cuts applied to the library. In such cases we exclude the candidate galaxy from the library, so that only suitable galaxies are selected.

\subsubsection{Luminosity Adjustment}
\label{sec:lum_adjustment}
A complication in the Match-Up Pipeline is that the baseDC2
galaxies are more luminous than the ones present in the Galacticus
library. To remedy the luminosity mismatch at the bright end, the
magnitudes for both sets of galaxies are compressed at the brighter
end into a smaller and overlapping range:

\begin{equation}
	M'_r(M_r) =
    \begin{cases}
    	\alpha \tanh \big( (M_r-M_{r0})/\alpha \big) +M_{r0},& \text{if } M_r \leq M_{r0}\\
        Mr,& \text{Otherwise}\\
    \end{cases}
    \label{eq:mag_compressed}
\end{equation}
where $M_r$ and $M'_r$ are, respectively, the original and compressed $r$-band magnitudes, $M_{r0}$ is the threshold where the compression
starts, and $\alpha$ is the range of the compression. Luminosities brighter
than $M_{r0}$ are compressed into a range that is strictly less bright
than $M_{r0}-\alpha$. The compression effectively downweights the
luminosity matching requirements at the brighter end while keeping the
color matching requirement fixed. Once a matching library galaxy is
found, the luminosity in each bandpass is rescaled by the same
constant factor that forces the $r$-band luminosity for the library and
baseDC2 galaxy to agree by construction.

\subsubsection{Assigning a Match}
\label{subsec:match}

The matching between baseDC2 and the library is done by constructing
k-d trees for the library and querying these trees for the nearest
neighbors to each baseDC2 galaxy. Since the library galaxy properties
change with redshift, the k-d trees have to be reconstructed each
time the library is interpolated to a new redshift. It would be
computationally impractical to construct the k-d trees for each
individual redshift value in the baseDC2 lightcone catalog.  Instead, we use
sets of k-d trees at narrow redshift slices of the lightcone as described below.

\begin{figure}
  \centering
  \includegraphics[width=8cm]{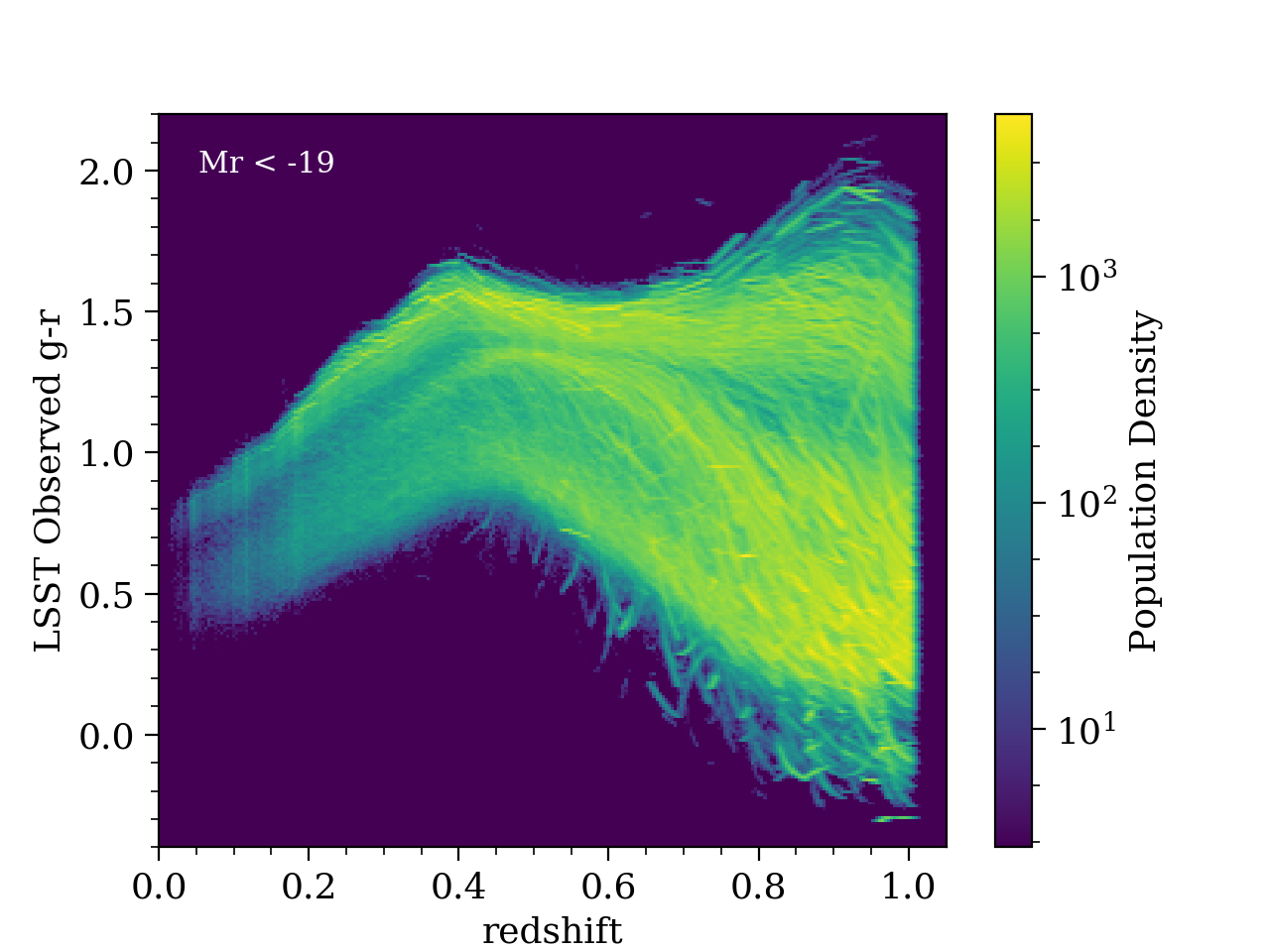}
  \caption{ Observed $g-r$ color distribution of cosmoDC2 galaxies with $M_{\rm r}<-19$ as a function of redshift,
    up to $z=1$. The smooth distribution is obtained through the
    interpolation procedure described in Sec.~\ref{subsec:match}. Filamentary
    structures visible in this figure arise from repeated sampling of library galaxies in
    sparsely populated color-magnitude space.}
  \label{fig:color-redshift}
\end{figure}

\begin{figure}
    \centering
    \includegraphics[width=0.45\textwidth]{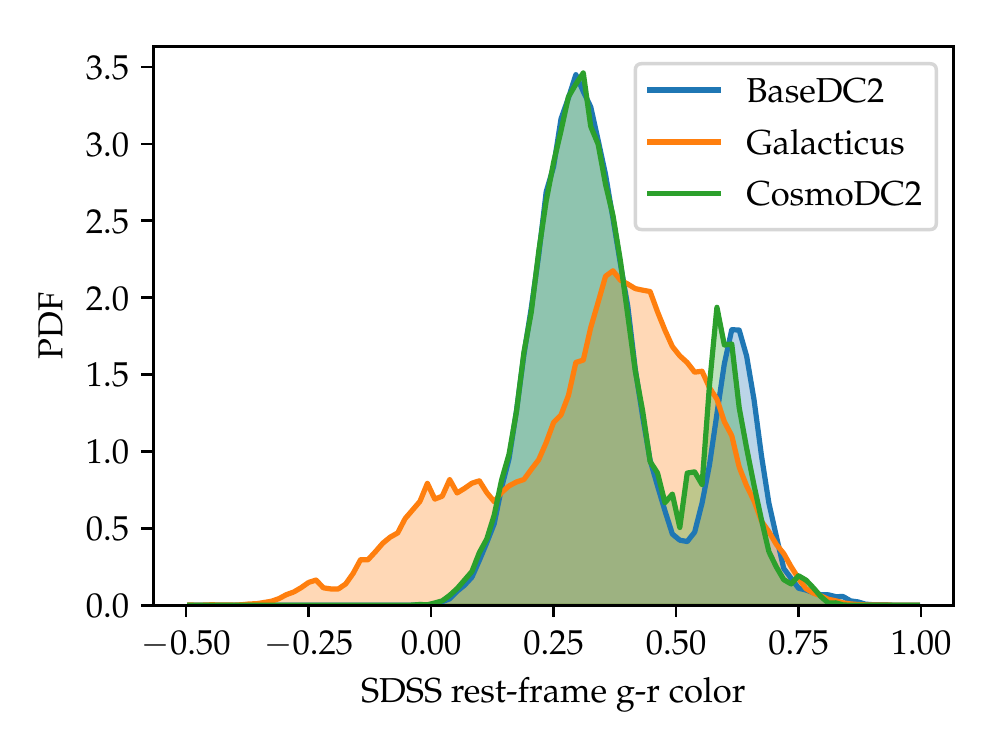}
    \centering
    \includegraphics[width=0.45\textwidth]{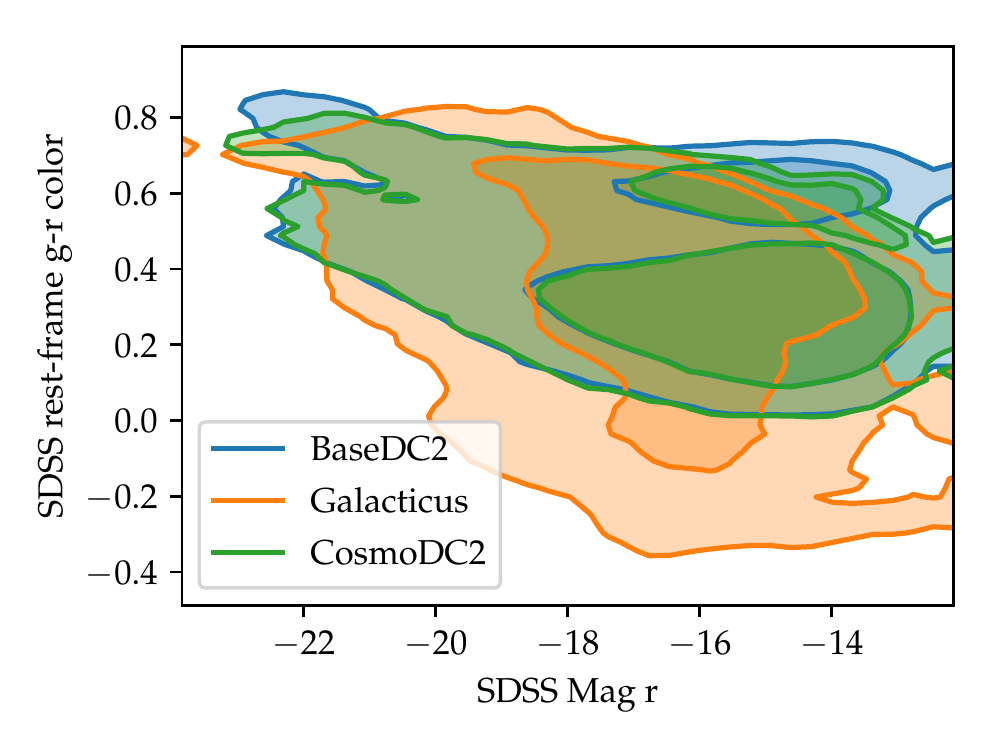}
  \caption{ Distributions of rest-frame color
    (top panel) and color-magnitude (bottom panel) of $0.5<z<0.54$ galaxies in baseDC2, the Galacticus library and cosmoDC2. The top panel shows that the color distribution of baseDC2 is well recovered in cosmoDC2 by selecting specific Galacticus galaxies, with only a small discrepancy for red galaxies. The contours shown in the bottom panel encompass 75\% and 99\% of the galaxy populations. 
    For bright galaxies, the luminosity adjustment prioritizes the match with baseDC2 colors while the luminosity rescaling extends the coverage of library galaxies (See Sec.~\ref{sec:lum_adjustment}). The brightest and reddest galaxies do not reproduce the baseDC2 colors exactly because there are few library galaxies with similar colors. The luminosity adjustment is applied for faint galaxies, so for faint red galaxies the match-up procedure  compromises galaxy color in order to match luminosity better.
    }
  \label{fig:match-color-distribution}
\end{figure}

We divide the redshift range between adjacent snapshots into five slices. For
each of these slices, the galaxy library is interpolated to its median redshift and two k-d trees are constructed from the library. The
first tree, which is used for most galaxies, only uses rest frame
$r$-band, $g-r$ and $r-i$ color and uses the distance metric of  Eq.~\ref{eq:matchup_dist}. The second tree, which is used for red sequence
cluster members, additionally uses observer frame $g-r$, $r-i$ and $i-z$
colors and uses the distance metric of Eq.~\ref{eq:matchup_dist_rs}. For each baseDC2 galaxy in the slice, we use the k-d tree
to find the ten closest neighbors in the galaxy library and assign a match
randomly from those ten, which smooths out discreteness effects otherwise caused by repeated selection of the same library galaxy in a sparsely populated area of color-magnitude space. Figure~\ref{fig:color-redshift} shows the color-redshift evolution of galaxies resulting from our match-up procedure.
The color distribution before and after the match-up procedure is shown in Figure~\ref{fig:match-color-distribution}. 

\subsection{Additional Modeling}

The final code module of Figure~\ref{fig:matchup-pipeline} shows the additional empirical modeling performed after the Galacticus library match-up. This modeling, which is required to meet validation criteria, relies on information obtained from the match-up.
Galaxy profiles are assumed to be given by an $n=1$ Sersic (exponential) profile for the disk component of each galaxy and an $n=4$ Sersic profile for the bulge component.

\subsubsection{Disk and Bulge Size}
\label{subsec:size}

The stellar mass of each galaxy in the Galacticus library is divided into disk and bulge components, from which we determine ${\rm B/T},$ the ratio of the stellar mass in the bulge to the total stellar mass. We use this quantity to model galaxy sizes in order to meet DESCQA validation criteria for the size-luminosity relation.

We separately model the size-luminosity relation for disk and bulge components, in both cases using the functional form introduced in \citet{zhang_wang17} to obtain $R_{50},$ the Petrosian half-light radius:
\begin{equation}
\label{eq:size_model}
R_{50} = \gamma L^{\alpha}(1+L)^{\beta-\alpha}.
\end{equation}
Here $L=10^{-0.4(M_{r}-M_{r}^{0})},$ and $\gamma, \alpha, \beta,$ and $M_{r}^{0}$ are fitting parameters that depend on the classification scheme used to determine the subtype of the fitted galaxy sample. \citet{zhang_wang17} characterize the size-luminosity relation for several classification schemes including a morphological classification into either elliptical or spiral types and a classification that uses the $\rm{B/T}$ value. We choose the latter scheme to obtain values of the fit parameters in Eq.~\ref{eq:size_model}. Specifically, for the disk and bulge components of cosmoDC2 galaxies at $z=0,$ we use the parameters in Table~1 of \citet{zhang_wang17} pertaining to SDSS galaxies with ${\rm B/T}<0.5$ and ${\rm B/T}>0.5$, respectively.

We incorporate redshift dependence in the galaxy-size distribution by setting $\gamma = \gamma(z)$. We choose the functional form of $\gamma(z)$ to be a sigmoid function, as defined by Eq.~\ref{eq:sigmoid}, which regulates the redshift evolution such that size decreases with increasing redshift. The parameters in Eq.~\ref{eq:sigmoid} are chosen such that we recover the \citet{zhang_wang17} parameters at $z=0$, the sizes of both disks and bulges are reduced by a factor of two at $z=1$ and $k$ is set to 4.

\subsubsection{Black Hole Mass and Accretion Rate}

A black hole resides at the center of every cosmoDC2 galaxy. The properties of these black holes can be used to model time-varying active galactic nuclei.  For the mass of the black hole, $M_{\bullet},$ we adopt the power-law scaling relation reported in \citet{kormendy_ho13}:
\begin{equation}
\label{eq:bhmass}
M_{\bullet} = 0.0049M_{\rm bulge}\left(M_{\rm bulge}/M_0\right)^{\alpha},
\end{equation}
where $\alpha=0.15$ and $M_0=10^{11}\rm{M}_{\odot}.$

For the mass accretion rate of the black hole, ${\rm d{\it M}_{\bullet}/dt},$ we define ${\rm dlog{\it M}_{\bullet}/dt}\equiv \lambda_{\rm edd}{\rm \dot{\it M}_{edd}},$ where $\lambda_{\rm edd}$ is the Eddington ratio and we assume an Eddington rate of ${\rm \dot{\it M}_{edd}}=0.022\rm{M}_{\odot}/{\rm Myr}/{\it M}_{\bullet}.$ We model $\lambda_{\rm edd}$ according to the redshift-dependent probability distribution reported in  \cite{aird_etal17a}:
\begin{equation}
\label{eq:bhaccretion}
P(\lambda_{\rm edd}\vert z) = A\frac{1+z}{(1+z_0)^{\gamma_{\rm z}}}\lambda_{\rm edd}^{\gamma_{\rm e}},
\end{equation}
where $A=0.00071,$ $\gamma_{\rm z}=3.47,$ $\gamma_{\rm e}=-0.65,$ and $z_0=0.6.$ Thus the specific mass accretion rate in this model has no dependence upon black hole mass, though ${\rm d{\it M}_{\bullet}/dt}\propto M_{\bullet}.$

In addition to the dependence of ${\rm d{\it M}_{\bullet}/dt}$ on black hole mass and redshift, we use conditional abundance matching (CAM) to introduce correlations between ${\rm dlog{\it M}_{\bullet}/dt}$ and ${\rm sSFR},$ the specific star-formation rate of the galaxy. For each galaxy, we calculate the cumulative probability ${\rm P(< sSFR\vert\mstar)},$ and use the CAM implementation in Halotools~\citep{2017AJ....154..190H} to non-parametrically correlate ${\rm sSFR}$ and $\lambda_{\rm edd},$ setting the correlation strength to $50\%.$ In cosmoDC2, galaxy ${\rm sSFR}$ is tightly correlated with broadband color, such that active galaxies have bluer colors; thus our use of CAM in assigning $\lambda_{\rm edd}$ produces synthetic catalogs in which galaxies with bluer broadband color host black holes that tend to be rapidly accreting mass.

\subsubsection{Galaxy Ellipticity}
The magnitude of the ellipticity of the galaxy is calculated as the $r$-band
luminosity-weighted average of the disk and bulge ellipticities.  These, in turn, are drawn from a Johnson SB probability distribution:
\begin{equation}
  \label{eq:ellip_model}
  f(e, a, b) = \frac{b}{e(1-e)}\phi\left(a+b\log\frac{e}{1-e}\right),
\end{equation}
where $e$ is the ellipticity, $\phi$ is the normal probability
distribution and $a$ and $b$ are model parameters. For the disk
ellipticity we use constant values of $a=-0.4$ and $b=0.7$. For the bulge component, we set $b=1.0,$ and $a$ to a value that depends on the rest frame $r$-band magnitude as follows: $a=0.6$ for $\magr \leq -21$, $a$ increases linearly with $\magr$ for $-21 \leq \magr \leq -17$ with slopes such that $a=1.0$ for $\magr=-19$ and $a=1.6$ for $\magr \geq -17$.  The values of these parameters have been chosen to match the ellipticity distributions reported in \cite{Joachimi2013}, who studied the shapes of galaxy images from the COSMOS survey.  The position angles, and thereby the components of the ellipticities, are chosen to correspond to galaxies with random orientations.

\subsection{Galaxy Catalog Content}

The catalog contains $\sim$2.26 billion galaxies in
a~440 deg$^2$ field that spans $0<z<3.$  Each galaxy
has 551 listed properties. The catalog size is 5.2 TiB and the catalog is subdivided into 393 HDF5 files separated by redshift range and sky pixelization.
In order to cover the image-simulation area that was selected for DC2, the catalog is built on the Nside=32 healpixels that are required to cover the area specified by the following (RA, Dec) coordinate pairs (J2000): (71.46, $-$27.25), (52.25, $-$27.25), (73.79, $-$44.33), (49.42, $-$44.33).
A comprehensive list of properties and specifications
including relevant units is given in Appendix~\ref{sec:galaxy_app} in
Table~\ref{tab:catalogcontents}.

Briefly, the modeled properties of galaxies in the cosmoDC2 lightcone include lensed and unlensed positions;  stellar mass and black hole mass; a range of luminosity information, including broadband flux through both LSST and SDSS bands, as well as coarsely binned SEDs, fluxes supplied with and without dust, in observer and rest frames, with separate fluxes for each galaxy's disk and bulge component; shape information, including ellipticity, shear, magnification, convergence, and size; we also include information about the parent dark matter halo of each galaxy.

\section{Selected Validation Results}
\label{sec:results}

In this section, we present selected validation results from cosmoDC2. These results have been chosen to be representative of important aspects of the output catalog and include redshift, magnitude and color  properties of the galaxies, and tests of large-scale clustering and lensing distortions. Calibration of the model was driven by the competing demands of DC2 for model complexity and accuracy, codified by the set of validation requirements supplied by DESC science working groups. The results shown here are a subset of the full range of DESCQA validation tests (to be presented elsewhere) which encompass additional tests that were used to validate cosmoDC2.

\subsection{Cumulative Number Counts as a Function of Magnitude}
\label{subsec:dndmag}

We compare the number counts of galaxies as a function of apparent magnitude with observational data.  These provide an important empirical test of the realism of the model for the redshift-dependent luminosity function.

\begin{figure}[!ht]
  \centering \includegraphics[width=8.5cm]{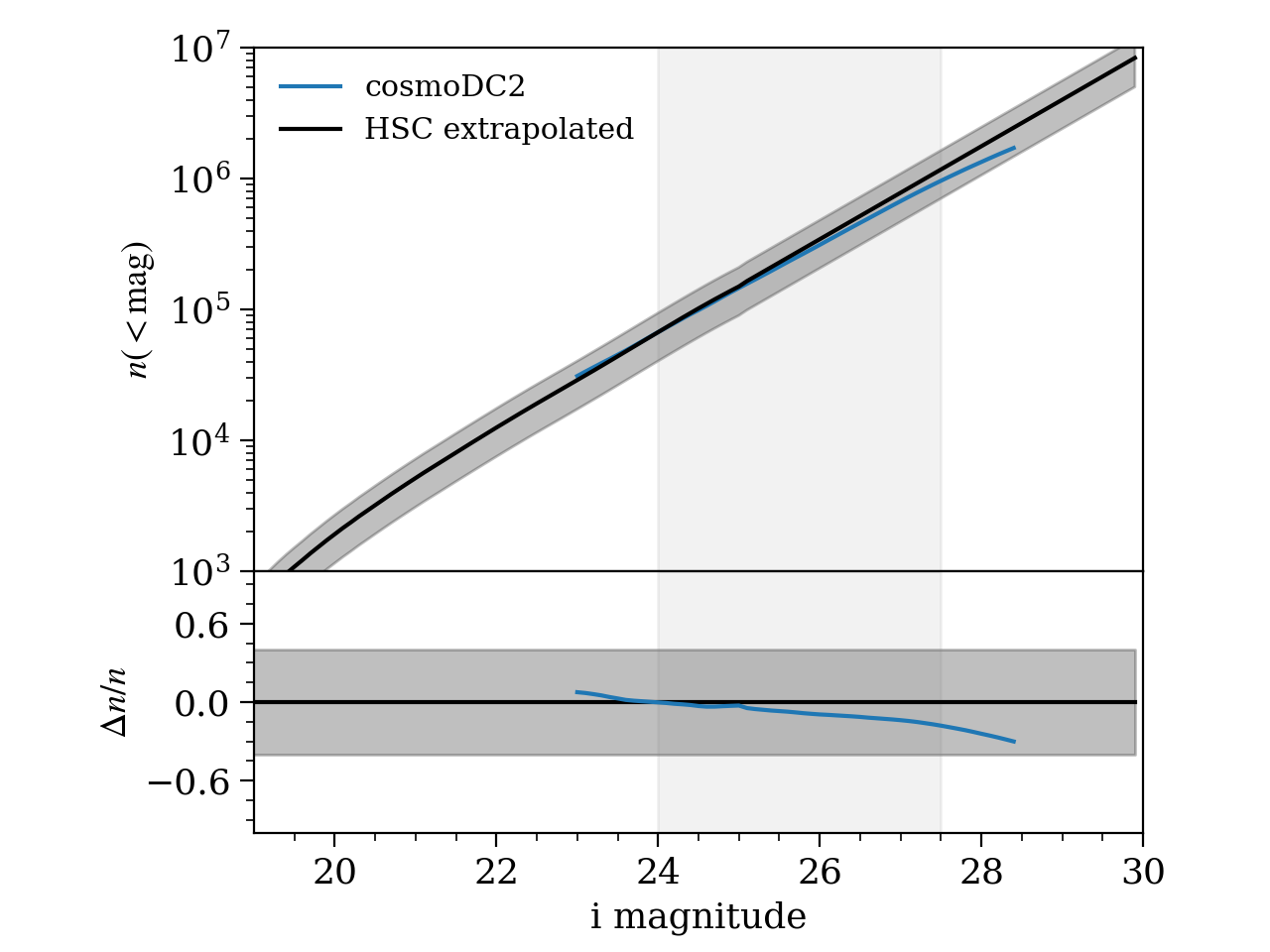}
  \caption{Top panel: Observed cumulative $i$-band number counts per square degree as a function of magnitude from cosmoDC2 (blue) and extrapolated from the HSC survey (black) (see text for more details). The grey shaded band shows a $\pm 40\%$ uncertainty around the HSC extrapolation. The vertical shaded region shows the magnitude range  within which the two curves are compared. Bottom panel: Relative difference between the two curves in the top panel. \label{fig:dndmag}}
\end{figure}

The dataset that we use for this comparison is the first data release \citep{2018PASJ...70S...8A} of the Deep layer of the Hyper-SuprimeCam (HSC) survey \citep{2018PASJ...70S...4A}. The reason for this choice is that the Deep layer of the HSC survey is the deepest ongoing survey with an area exceeding a few square degrees. Hence, we perform a comparison that involves less extrapolation than with a shallower large-area survey, but without the cosmic variance uncertainty that would come from using a much smaller pencil beam survey. Using this data set, we measure the cumulative number counts of galaxies in the $i$-band down to 25th magnitude, then fit the result to a power law. This power-law extrapolation is justified by measurements from deep HST surveys, which appear to have power-law number counts down to at least 28th CModel magnitude \citep[e.g.,][]{2006AJ....132.1729B}. The results are shown in \autoref{fig:dndmag}.

While this plot was used for validation, additional checks were made of the number counts against other surveys.  For example, Subaru observations in the COSMOS field \citep{2007ApJS..172...99C} yield a cumulative number density of 150 arcmin$^{-2}$ or $5.4\times10^5$ deg$^{-2}$ for $I<26.5$, which is clearly quite close to the extrapolated HSC curve in \autoref{fig:dndmag}.

The grey-shaded band around the HSC extrapolated curve shows the region of $\pm 40\%$ tolerance in the cumulative number counts, the validation criterion set by the DESC science working groups. The magnitude range for validation region is indicated by the vertical shaded band which covers the range $24 < r < 27.5$. The maximum fractional difference between the cosmoDC2 data and the HSC fit in this region is 0.17 and hence cosmoDC2 passes this important validation test. As discussed in Sec.~\ref{subsec:faint}, the decrease in the number count with increasing $i$-band magnitude is an effect of the mass resolution of the simulation. Before the inclusion of the ultra-faint galaxies described in Sec.~\ref{subsec:faint}, cosmoDC2 failed this test.

\subsection{Redshift Distribution}
\label{sec:reddist}

\begin{figure}
  \centering \includegraphics[width=8.5cm]{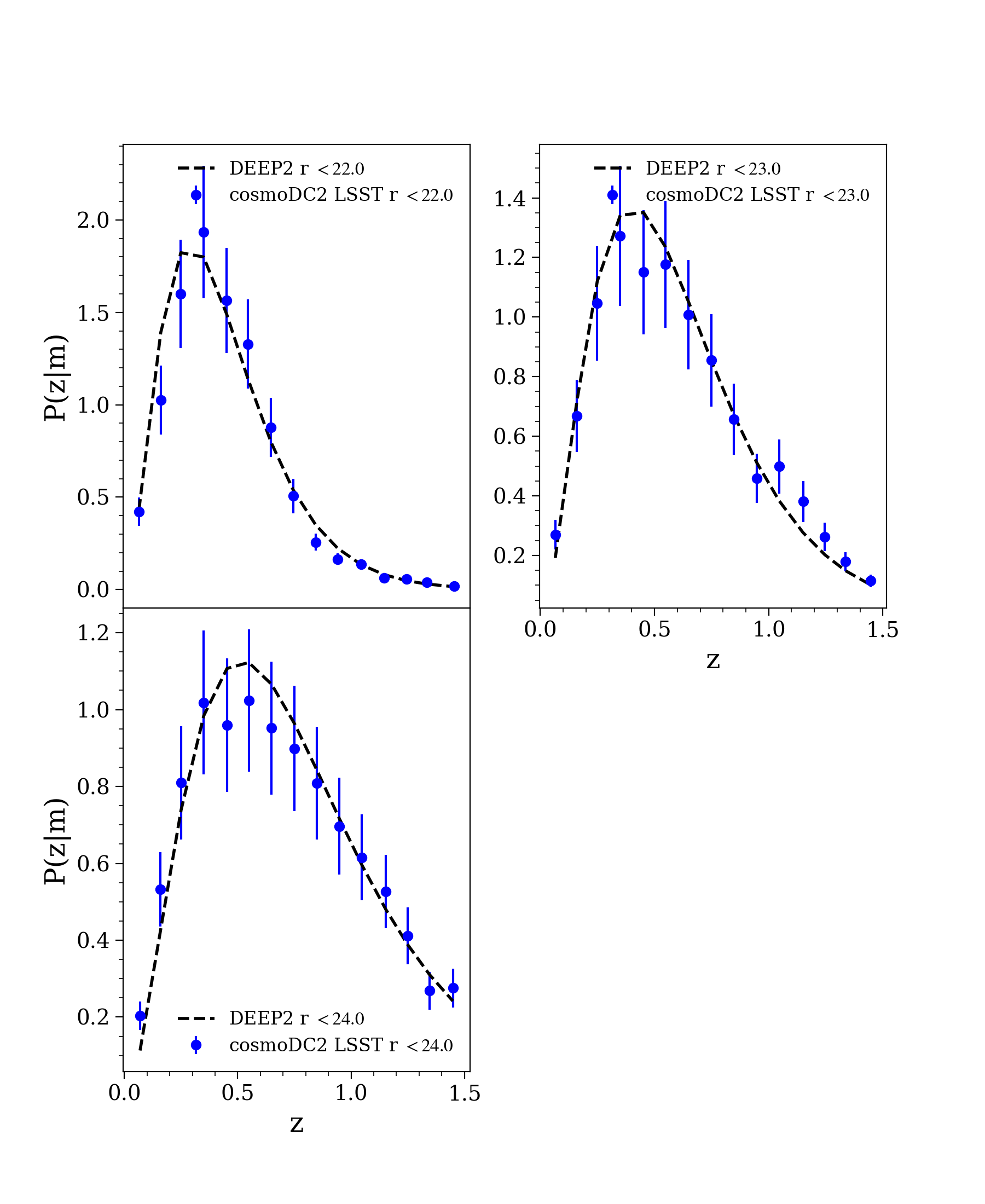}
  \caption{Redshift distribution of cosmoDC2 galaxies compared with fits to DEEP2 data for  a redshift range of $0 < z <1.5$ for three magnitude-limited samples. The selection cuts are LSST-$r<22.0$, $r<23.0$ and $r<24.0$ as indicated in the legend.}
  \label{fig:nz}
\end{figure}

The distribution of the number of galaxies as a function of redshift is another fundamental test of the realism of the synthetic catalog. This test is complementary to the cumulative-number-density test described above and provides a check on the shape of the redshift-dependent luminosity function. For this test, we compare the probability distribution for the number of galaxies as a function of the cosmological redshift with the observational data from \cite{deep2-dndz} and \cite{coil}. These observations are reported as parameterized fits to the d$N$/d$z$ distributions for a variety of magnitude-limited samples. The selection cuts for these samples were imposed on the CFHT $R$- and $I$-band magnitudes.  In order to obtain comparisons that are as meaningful as possible, we construct magnitude-limited samples for the catalog data by applying the same cuts to the $r$- and $i$-band LSST filter magnitudes in the cosmoDC2 data. The error $\sigma_i$ for each redshift bin is determined with a jackknife procedure: we estimate the errors due to cosmic variance by excluding subregions of RA and Dec from the catalog footprint with a k-means algorithm as implemented in the scikit-learn package.\footnote{\url{https://scikit-learn.org/stable/}}. The elements of the covariance matrix are then given by
\begin{equation}
    \sigma^2_{ij} = \frac{N_{\rm{jack}}-1}{N_{\rm{jack}}} \sum_k (\bar{N}_i - N^k_i)\cdot(\bar{N}_j - N^k_j) ,
\end{equation}
where $N_{\rm{jack}}$ denotes the number of jackknife regions (chosen to be 30 for the present work) and $\bar{N}_i$ and $N^k_i$ denote the numbers of galaxies in the $i$-th redshift bin for the full sample and for the sample with the $k$-th region excluded, respectively. The score for the test is computed by calculating the average of the reduced $\chi^2$ between the catalog data and observed fit for each of the magnitude-limited samples. In practice, the computation of the covariances can be lengthy and we typically run this test over smaller sky areas of $\sim$ 100 deg$^2$ at a time.  For these smaller areas, the above covariance matrix is often not invertible due to instabilities in the off-diagonal matrix elements, so we use only the diagonal elements in the $\chi^2$ computation. In Figure~\ref{fig:nz}, we compare the redshift distributions for cosmoDC2 for three magnitude-limited samples having a sky area of $\sim$ 60 deg$^2$ with the fits obtained from the DEEP2 data. The redshift distributions are in reasonable agreement with the DEEP2 fits, and we obtain $\chi^2/\rm{d.o.f}$ values of 1.2, 0.75 and 0.73 for magnitude cuts of $r<22$, $r<23$ and $r<24$, respectively. These values of $\chi^2/\rm{d.o.f}$ may be somewhat under-estimated due to the aforementioned problem of obtaining the off-diagonal elements of the covariance matrix. We note that the validation criteria for cosmoDC2 do not specify a quantitative tolerance for $\chi^2/\rm{d.o.f}$.

\subsection{Color Distributions}
\label{subsec:color}

\begin{figure}
  \centering \includegraphics[width=9.0cm]{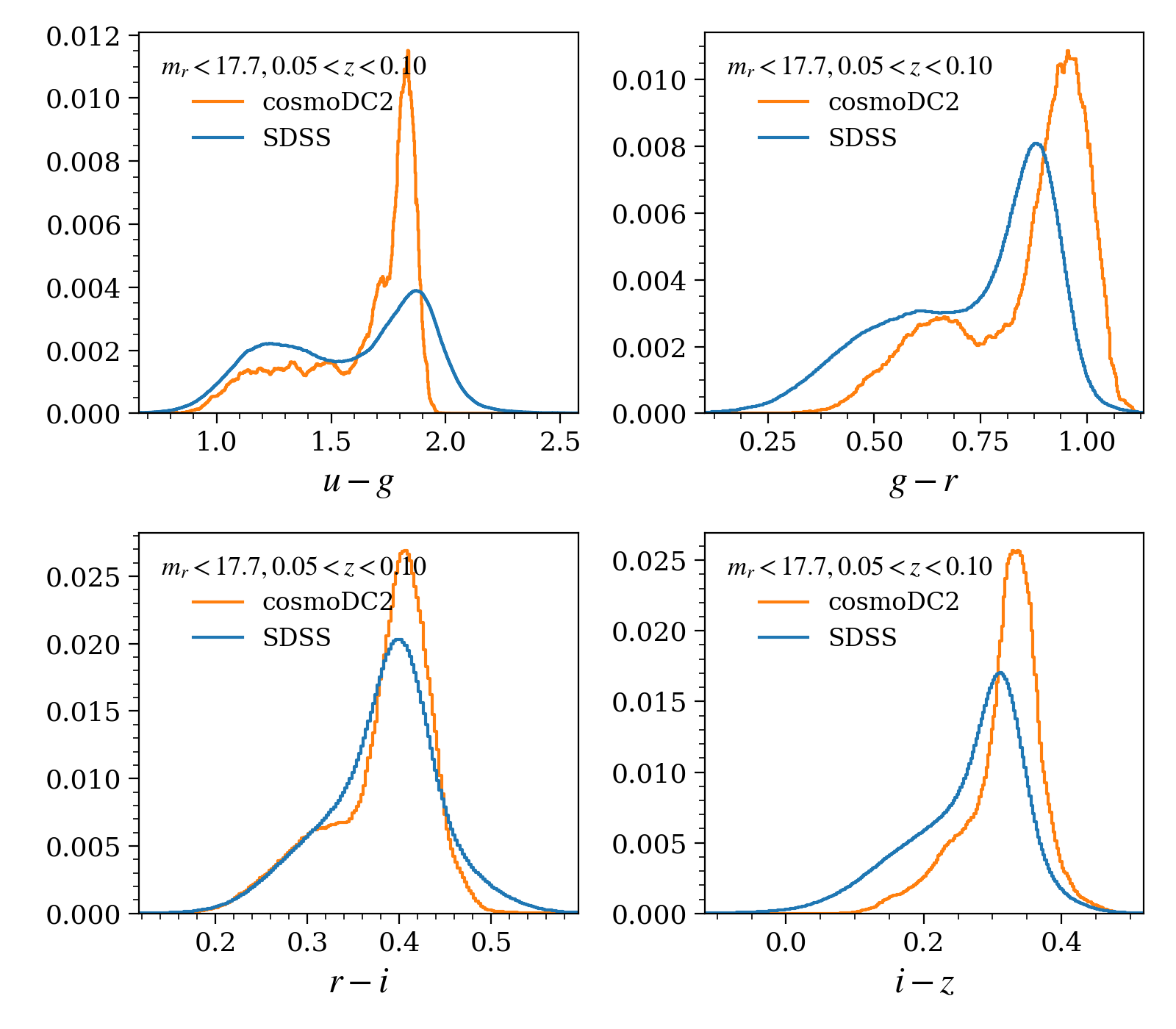}
  \caption{SDSS color distributions of cosmoDC2 galaxies compared with those of SDSS data. The cosmoDC2 galaxies comprise a magnitude- and redshift-limited sample with SDSS $r$-band magnitude $<17.7$ and a redshift range of $0.05 < z < 0.1$. The colors shown in the figure are SDSS $u-g$, $g-r$, $r-i$ and $i-z$.}
  \label{fig:color}
\end{figure}

The color distributions of galaxies as a function of redshift and luminosity are critical properties that must be rendered with sufficient realism for many of the DC2 scientific use cases. For example, the accurate characterization of the biases and systematics of photometric redshift algorithms relies on the realism of synthetic color distributions. In Figure~\ref{fig:color}, we compare the color distributions for a magnitude-limited sample of cosmoDC2 galaxies with SDSS $r$-band $< 17.7$ and redshifts in the range $0.05 < z < 0.1$ to observational data from the SDSS main galaxy sample in SDSS DR13 \citep{2017ApJS..233...25A}. The validation criteria for this test are that the bimodalities, peak locations and luminosity dependencies of the cosmoDC2 color distributions are broadly in agreement with those of the SDSS data.

Two features of the comparisons between cosmoDC2 and the SDSS data are worth noting. First, recall that the empirical model is tuned only for {\it rest-frame} $r$-band magnitude and $g-r$ and $r-i$ colors, whereas the quantities shown in the figure are {\it observer-frame} colors. The level of agreement that has been achieved relies solely on the correlations between the properties available for tuning and the full set of galaxy properties, as discussed in Sec.~\ref{sec:e-to-e}. Second, the distributions for cosmoDC2 peak at redder colors for $g-r$ and $i-z$ distributions and a bluer color for $u-g$. There are no empirical parameters available to alter $u-g$ and $i-z$ color distributions independently of the others, so it is difficult to achieve better agreement for the joint distribution of all colors by changing the empirical model. Other constraints, such as the requirement that the catalog have a prominent sample of red-sequence galaxies, impose further restrictions on the joint color distributions, and so the comparison shown here represents a compromise between multiple criteria supplied by the DESC science working groups.

\subsection{Two-Point Correlation Function}

Galaxy clustering measurements provide a biased, but high signal-to-noise probe of the underlying matter field, and are commonly used both individually and in combination with lensing measurements to constrain the underlying matter field \citep[e.g.,][]{des32pt}. We therefore validate the catalog for such measurements by comparing mock observations of the galaxy two-point correlation function to data. In this highlighted test we use the \texttt{TreeCorr} package \citep{treecorr} to mimic measurements from \cite{tpcf_wang} of $w(\theta)$, the over-abundance of galaxy pairs at some angular separation $\theta$ relative to a random distribution,  given selected SDSS $r$-band magnitude cuts. This uses the estimator of \citet{landy_szalay}
\begin{equation}
    w(\theta) = \frac{N_{dd}-2N_{dr}+N_{rr}}{N_{rr}},
\end{equation}
where $N$ is a normalized pair count within an angular separation bin centered at $\theta$, and the subscripts $d$ and $r$ refer to the data or randomly generated galaxies used to compute pairs, respectively.

\begin{figure}[!ht]
  \centering
  \includegraphics[width=8.75cm]{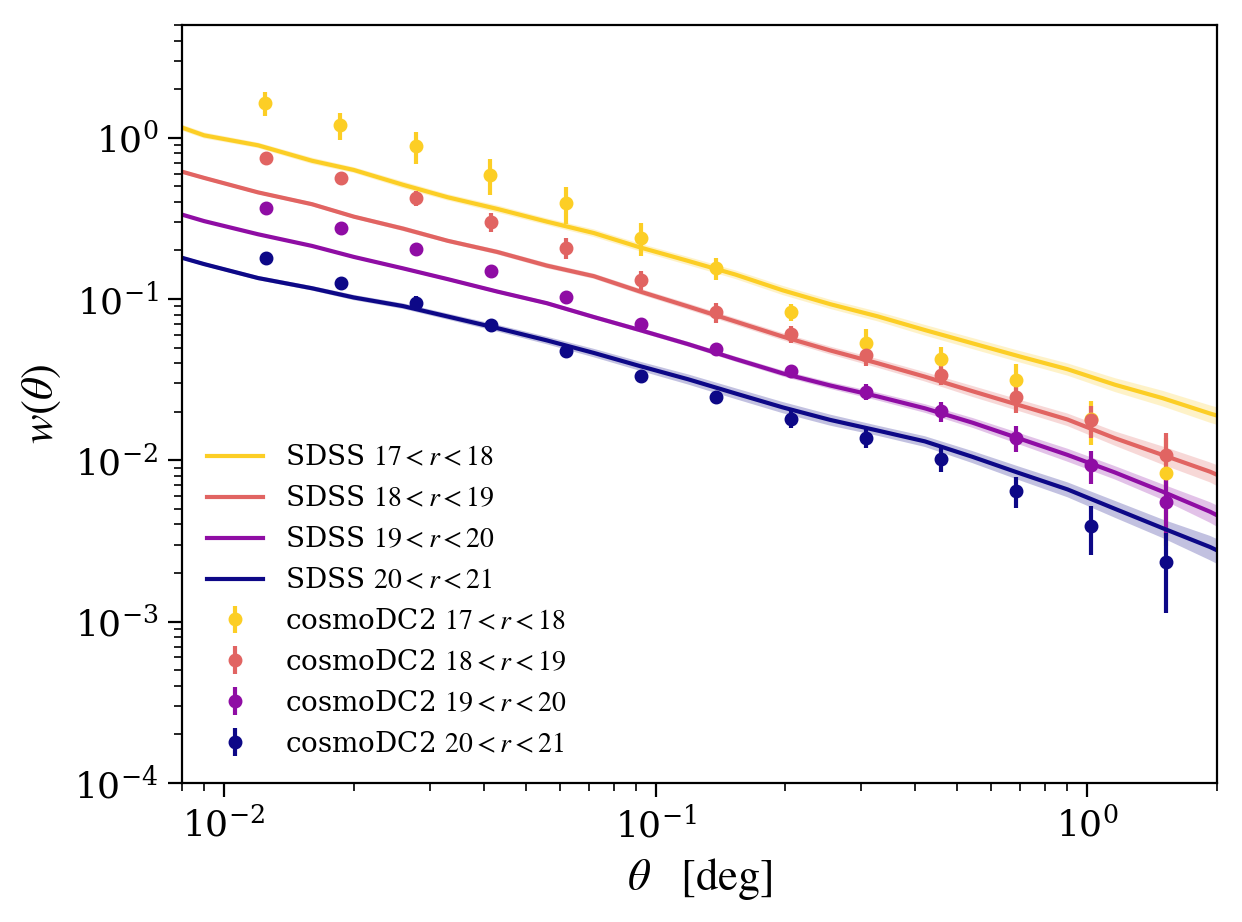}
  \caption{Validation test for the two-point correlation function, computed on the 440 sq. deg. cosmoDC2 catalog for cuts in SDSS $r$-band magnitude as given in the legend. The points correspond to measurements on the cosmoDC2 catalog, with error bars obtained through jackknife resampling,  and solid
    lines to SDSS measurements from Table 2 of \citet{tpcf_wang}.}
  \label{fig:2pt}
\end{figure}

Figure \ref{fig:2pt} shows the shape and amplitude of the angular clustering of different galaxy samples selected by apparent $r$-band magnitude. The overall trends with scale, amplitude, and luminosity are the same as seen in SDSS; more pronounced deviations are visible in the 1-halo term for brighter galaxy samples, particularly the steeper slope of the clustering in the mock relative to SDSS. To estimate the variance within this patch we have included error bars using the jackknife procedure detailed in Sec.~\ref{sec:reddist} with $N_{\rm{jack}}=20$.

For this test, the validation criteria supplied by the DESC science working groups for the purposes of DC2 amounted to a check that the synthetic catalog and validation data were in reasonable agreement, and that the clustering strength scales with galaxy brightness in the expected fashion. Achieving higher precision agreement in the future would involve a much more costly optimization exercise, as well as a larger catalog area to reduce the effect of cosmic variance, and a more realistic estimation of statistical and systematic errors.

\subsection{Galaxy-galaxy Lensing}

\begin{figure}[!ht]
  \centering \includegraphics[width=8.5cm]{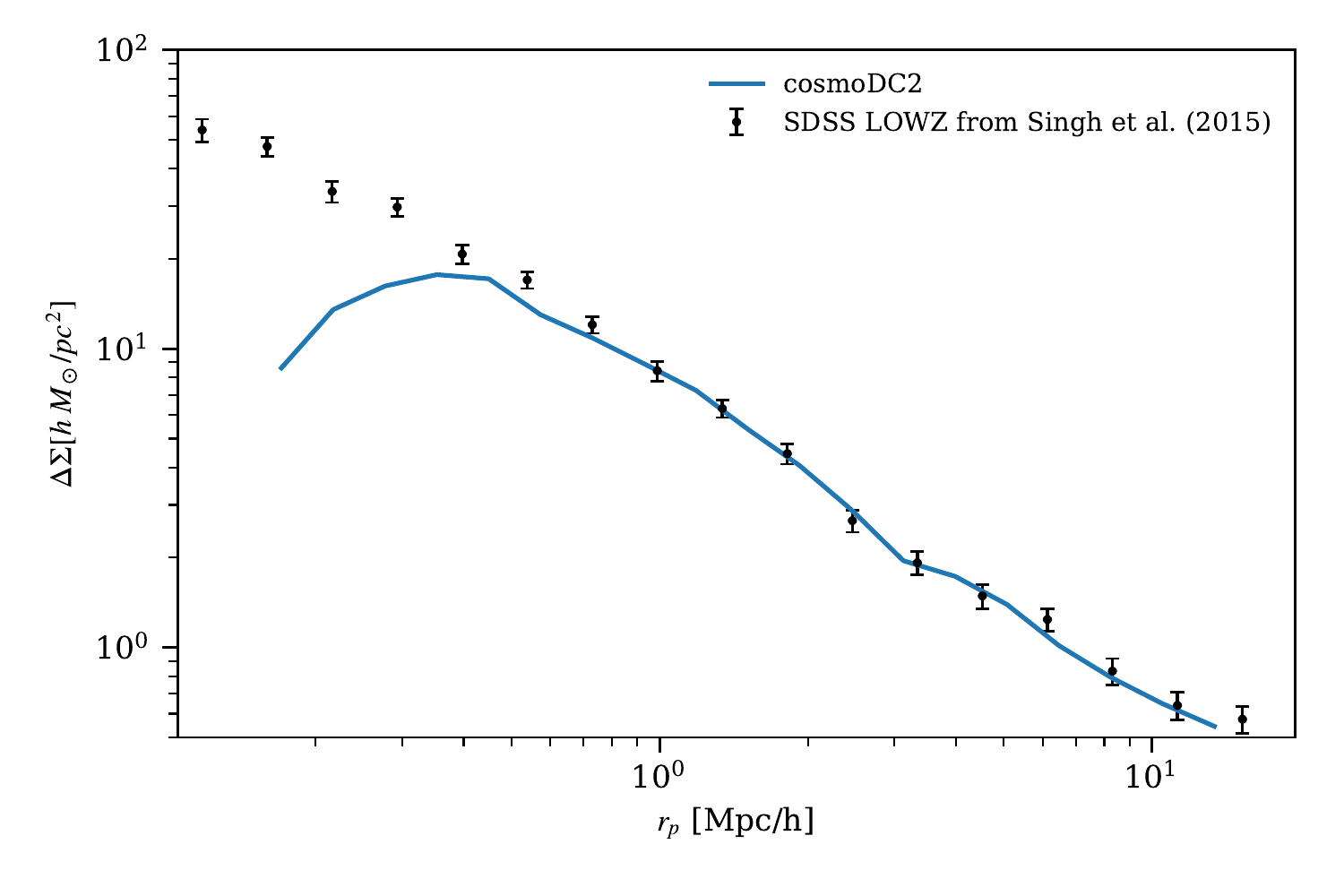}
  \caption{Validation test for galaxy-galaxy lensing, computed on the cosmoDC2 image area with cuts chosen to match those of the SDSS LOWZ sample of \citet{2015MNRAS.450.2195S}. The test returns a total SDSS LOWZ galaxy number density of 58.8 per square degree for the cosmoDC2 image area, compared to the observed value of 57.0 per square degree from \citet{2016MNRAS.455.1553R}. The solid blue line in the figure corresponds to the measurement performed on the cosmoDC2 catalog, and the black points to the measurement on the SDSS sample. At small scales the signal is smoothed due to lensing map resolution limits.}  \label{fig:dst}
\end{figure}

Figure~\ref{fig:dst} displays the results of the galaxy-shear correlation test 
on the cosmoDC2 catalog. This computes the average tangential shear $\gamma_t$ of a collection of background source galaxies at a given projected physical distance $R$ from foreground lens galaxies, where
\begin{equation}
\gamma_{t} = -[\gamma_{1}\cos(2 \phi_{c})+\gamma_{2}\sin(2 \phi_{c})].
\end{equation}
Here $\phi_{c}$ is the angle of the vector connecting the projected lens and source galaxies. These values are scaled by the geometry-dependent critical surface density $\Sigma_{\rm{crit}}$ to give the excess surface mass density $\Delta\Sigma (R) $. This test uses color, magnitude and redshift cuts designed to mimic the SDSS LOWZ measurement of \citet{2015MNRAS.450.2195S}, compute the excess surface mass density, and compare it to the observed values.

While no quantitative validation criteria were provided by DESC science working groups for this test, nonetheless the cosmoDC2 results show a qualitatively good fit to the SDSS data at large scales, and a realistic galaxy number density for the LOWZ sample; the falloff of the synthetic lensing signal on small scales is expected due to pixelization and noise in the shear maps as noted in Sec.~\ref{sec:shear}. This test shows that the LOWZ-like population in the catalog is very similar to observations, both in number density and correlation with the underlying tangential shear field, validating an important use case of the catalog, as well as confirming that the weak-lensing quantities and galaxy positions are appropriately correlated.

\section{Summary and Future Directions}
\label{sec:outlook}

In this paper, we have described the production of cosmoDC2, a large synthetic sky catalog tailored to the needs of contemporary cosmological surveys. CosmoDC2 serves as the extragalactic catalog used in the end-to-end image simulation pipeline developed as part of the LSST DESC Data Challenge 2. To produce the synthetic data, we have developed a new methodology for modeling the galaxy-halo connection that combines empirical and semi-analytic models (A. Hearin et al. 2019, in preparation), as well as a new software pipeline for ray-tracing computations (P. Larsen et al. 2019, in preparation). The cosmoDC2 lightcone spans the redshift range $0<z<3;$ each galaxy in cosmoDC2 has more than 500 attributes, including broadband flux through LSST filters, stellar mass, gravitational shear, separate coarse-grained SEDs for disk and bulge components of the image, half-light radius, black hole properties, and a range of other properties. The requirements for the statistical distributions of galaxies in cosmoDC2 were designed in close collaboration with the analysis working groups in DESC, using the DESCQA validation software to ensure appropriate realism.

Here, we have given a detailed account of cosmoDC2, the extragalactic catalog used in the DC2 image simulation, which spans $440\ {\rm deg}^2.$ In the near future, we will use our pipeline to generate a $5000\ {\rm deg}^2$ synthetic sky; this larger catalog will be used in a range of scientific analyses conducted by DESC science working groups.

Our effort to produce high-quality synthetic data for LSST DESC is ongoing, and several specific areas have already been targeted for further improvement. For example, the explicit halo-mass dependence of the SEDs of galaxies in the red sequence in cosmoDC2 galaxies is inconsistent with assumptions made by RedMaPPer. The additional observer-frame color matching that is done for red-sequence cluster members introduces a difference between cosmoDC2 red-sequence field and cluster galaxies. Recalibrating the color model after eliminating this explicit halo-mass dependence will improve the applicability of the catalog for studies of redMaPPer projection effects.

We also plan to improve the physical realism of the spatial distribution of cluster satellites. The current intra-halo distribution has a truncated Navarro-Frenk-White (NFW) profile~\citep{1996ApJ...462..563N} that is inherited from Rockstar subhalos that are populated with UniverseMachine galaxies; in the future, the radial profiles will no longer be truncated and will follow an ellipsoidal NFW profile that is aligned with the large-scale tidal field.

In future releases of the model, we plan to extend our framework to utilize hydrodynamical simulations. For example, gas profiles from high-resolution hydrodynamical simulations can be painted onto group and cluster halos using the Galsampler technique, creating mock catalogs that could be used to study baryonic effects in the environments of massive halos.

Finally, we will soon provide high-resolution cutouts of the density and shear field surrounding cluster-mass halos. A comprehensive and up-to-date list of all planned improvements is available.\footnote{\url{https://github.com/LSSTDESC/cosmodc2/issues}}

Due to the evolving nature of the validation criteria of modern surveys, any method for generating synthetic cosmological data must be flexible enough to accommodate the demands imposed upon it by validation, as well as have sufficient computational efficiency to facilitate repeated iteration. Our hybrid method has a number of features that make it particularly suitable for producing a catalog that meets a series of potentially evolving validation criteria. The workflow is quite flexible in that the empirical models that drive the initial distributions of the limited set of galaxy properties can be easily changed. The pipelines can be run relatively quickly so that it is quite feasible to iterate on the empirical model to improve the agreement between the catalog and the observational data. The most  time-consuming part of the modeling, namely running the SAM, need only be done once. If the resulting library spans the range of observed properties, the matching with empirically-modeled galaxies is straightforward. In the future, as more observational data become available from deeper surveys, the validation of synthetic catalogs will become much more demanding. It will be crucial to continue development of methodologies that efficiently and flexibly generate realistically complex synthetic cosmological data.

The cosmoDC2 catalog is publicly available from the NERSC website\footnote{See \url{https://portal.nersc.gov/project/lsst/cosmoDC2/_README.html}} as a collection of HDF5 files. The files are labeled by redshift range and healpixel numbers, which correspond to the Nside = 32 healpixels in the DC2 image-simulation area. As described in Appendix \ref{sec:reader}, the LSST~DESC provides a Python package to facilitate user access to cosmoDC2.
The cosmoDC2 code is publicly available.\footnote{ \url{https://github.com/LSSTDESC/cosmodc2}}

\section*{Acknowledgements}
\label{sec:acknowledgements}

This paper has undergone internal review by the LSST Dark Energy Science Collaboration. The internal reviewers were Matt Becker, Alexie Leauthaud, and Nelson Padilla. We thank Seth Digel for his careful reading of the manuscript and his thoughtful comments. 

DK led the match up between the baseDC2 and Galacticus pipeline and was closely involved in many aspects of the catalog production.
APH helped devise the underlying model for the galaxy-halo connection, wrote the GalSampler package, and guided development of the pipeline.
EK ran the Galacticus simulations, developed code for and ran the baseDC2 production pipeline, and worked on many aspects of the validation of the catalog.
PL developed the full-sky version of the lensing pipeline, and contributed to various production and validation efforts.
ER made contributions to the text of the paper and to the codes used for lightcones and building merger trees.
JH developed the interpolation module for lightcone construction, and assisted in catalog validation and writing of the paper text.
AJB developed the Galacticus code and assisted in running it on the simulations underlying this paper.
KH carried out the N-body simulations underlying this work and also performed the first level analysis of the simulations (halo finding etc.) She has contributed to the text of the paper and was engaged throughout in all aspects of the project.
YYM contributed to the catalog readers, validation of the catalogs, and the text of the paper.
AB tested the galaxy stellar mass distribution against SDSS BOSS and tested the galaxy number density.
CC worked on the galaxy-galaxy lensing tests with different lens samples.
DC contributed to the catalog readers and validation of the catalogs.
JD consulted on the galaxy-halo connection in the cluster regime and contributed validation tests for cluster populations and galaxy color distributions.
HF is one of the key HACC developers and contributed to the simulations underlying this paper. He wrote an early extrapolation method for lightcone construction upon which the work presented here was based.
NF is an important member of the HACC team and developed several of the tools used for the analysis of the N-body simulation results.
EG participated in validation test definition and provided detailed feedback on a draft of the manuscript.  
SH is the HACC team lead and contributed to the simulations and methodological development underlying this project. He was engaged in developing the synthetic galaxy catalog concepts and also contributed to the text of the paper.
BJ contributed the COSMOS data that were used to model the galaxy ellipticity distributions.
FL contributed to the galaxy-galaxy lensing delta sigma and number density validation test.
NL worked on the multiple-lens-planes/source-planes ray-tracing simulation to assign weak lensing signals (including shears, convergence, and magnifications) to the galaxies in DC2.
RM provided high-level input and coordination regarding extragalactic catalog needs for all DESC science cases, and participated in validation test definition for several science cases (WL, LSS, PZ).
CM created validation tests comparing the correlation functions of stellar mass selected samples at z=1.0 to those measured by DEEP2.
JAN contributed various ideas for and assessments of tests versus real data sets, and provided DEEP2 redshift distributions for comparisons.
AP is a core developer of HACC and contributed to the simulations underlying this paper, and also assisted with early efforts to run Galacticus on HACC outputs.
ESR provided feedback on red-sequence cluster members, including tests and validation of color and scatter as a function of redshift.
MS contributed to the validation of the catalogs for weak lensing science.
CHT contributed to the validation of cosmoDC2 by running RedmaPPer on the catalog and by investigating the behavior of red sequence.
VV worked on the relationship between the luminosity and size of galaxies.
RHW contributed to improving and validating the galaxy--halo connection in the cluster regime.
MW contributed to a number of pilot studies and the early configuration and running of Galacticus.

YYM is supported by the Samuel P.\ Langley PITT PACC Postdoctoral Fellowship.

The work of HF, NF, SH, AH, KH, JH, DK, EK, PL, and ER at Argonne National Laboratory was  supported under the U.S. DOE contract DE-AC02-06CH11357.

LSST~DESC acknowledges ongoing support from the Institut National de Physique Nucl\'eaire et de Physique des Particules in France; the Science \& Technology Facilities Council in the United Kingdom; and the Department of Energy, the National Science Foundation, and the LSST Corporation in the United States. LSST~DESC uses the resources of the IN2P3 Computing Center (CC-IN2P3--Lyon/Villeurbanne - France) funded by the Centre National de la Recherche Scientifique; the National Energy Research Scientific Computing Center, a DOE Office of Science User Facility supported by the Office of Science of the U.S.\ Department of Energy under contract No.\ DE-AC02-05CH11231; STFC DiRAC HPC Facilities, funded by UK BIS National E-infrastructure capital grants; and the UK particle physics grid, supported by the GridPP Collaboration.  This work was performed in part under DOE contract DE-AC02-76SF00515.

This research used resources of the Argonne Leadership Computing Facility, which is a DOE Office of Science User Facility supported under Contract DE-AC02-06CH11357.

Computational work for this paper was also performed on the Phoenix cluster at Argonne National Laboratory, jointly maintained by the Cosmological Physics and Advanced Computing (CPAC) group and by the Computing, Environment, and Life Sciences (CELS) Directorate.

A portion of this work was performed at the Aspen Center for Physics, which is supported by National Science Foundation grant PHY-1607611.

The data for the DEEP2 Galaxy Redshift Survey were obtained at the W.M. Keck Observatory, which is operated as a scientific partnership among the California Institute of Technology, the University of California and the National Aeronautics and Space Administration. The Observatory was made possible by the generous financial support of the W.M. Keck Foundation.  Funding for the DEEP2 Galaxy Redshift Survey was  provided by NSF grants AST-95-09298, AST-0071048, AST-0507428, and AST-0507483 as well as NASA LTSA grant NNG04GC89G.  Code used to produce DEEP2 redshift distributions was modified from code originally written by Alison Coil and used in \citet{coil_etal04}.

A portion of this research was carried out at the Jet Propulsion Laboratory.
RM is supported by the US Department of Energy Cosmic Frontier program, grant DE-SC0010118.

We thank the developers and maintainers of the following software used at various stages of our catalog-making pipeline: Python \citep{python_language},
IPython \citep{ipython},
Jupyter \citep{jupyter},
NumPy \citep{numpy},
SciPy \citep{scipy},
Cython \citep{cython},
h5py (\href{http://www.h5py.org}{h5py.org}),
Matplotlib \citep{matplotlib},
Astropy \citep{astropy},
Healpix \citep{healpix},
Lenspix \citep{lenspix},
Polspice \citep{polspice}, CAMB \citep{lewis_challinor06}.

\bibliographystyle{yahapj}
\bibliography{refs,software}

\input{appendix.tex}
\end{document}

%% file: appendix.tex
\appendix

\section{Lightcone Construction}
\label{sec:lc_appdx}
While simulation time snapshots may serve as a substrate for shallow survey catalogs, cosmoDC2 spans a redshift range over which there is significant evolution in galaxy properties, and in the growth of large-scale structure. Thus, the contents of the catalog at any particular redshift should be built upon the matter distribution of a corresponding  snapshot of its parent N-body run. 

Of course, the notion of an object  being associated with a certain redshift is only meaningful with respect to some observer, which we can represent at any arbitrary point within the simulation volume. Then, the previous paragraph can be equivalently stated as such:  the signal that an observer $A$ receives from a distant source should be one that was emitted at some event located on $A$'s past lightcone.  The collection of all such events are those that have a null spacetime-separation, $ds$, from $A$, given by the Robertson-Walker metric, which satisfy the expression 
\begin{align}\label{eq:lc_rw}
\nonumber ds^2 &= -c^2dt^2 + a(t)^2\left[ dr^2 + S_\kappa(r)^2d\Omega^2\right] = 0\\ 
&\implies \int_{t_e}^{t_0}c\frac{dt}{a(t)} = r,
\end{align}
where $S_\kappa(r)$ is the spacetime curvature, $d\Omega$ is the change to spherical coordinates, and $r$ is the comoving radial displacement of the event from the observer. All events whose spacetime coordinates satisfy this condition lie on observer $A$'s past lightcone at cosmic time $t_e$, and are seen today at $t_0$ ($a=1$).

In order to convert from the matter distributions of discrete simulation box time snapshots, to one that is     ``observed'' across a simulated sky, we have developed a lightcone generation code module to approximate $r$, and evaluate Eq.~\ref{eq:lc_rw} for $a(t_e)$, for any collection of input events. In our case, those inputs are either simulation particle or halo positions, along with their snapshot times (see Section~\ref{sec:lightcone}). The general result of this process is an all-sky catalog of objects populating a smooth redshift distribution, as viewed by an arbitrary observer. 

Despite pristine knowledge of particle positions determined by cosmological simulations, finding the lightcone-crossing times of a catalog of objects is complicated by the fact that we are limited in temporal resolution, and have only of order 100 time snapshot outputs from $a \approx 0$ to $a=1$. Therefore, the intersection of a particle's worldline with the past lightcone of some observer is an event that is only captured in the simulation output as follows: in the snapshot immediately preceding that event, the particle's separation from the observer is timelike ($ds^2 < 0$), and in the following snapshot, it is spacelike ($ds^2 > 0$). 

\begin{figure}[h]
  \hspace{0.1\linewidth}
  \begin{minipage}[c]{0.42\linewidth}
    \includegraphics[width=\textwidth]{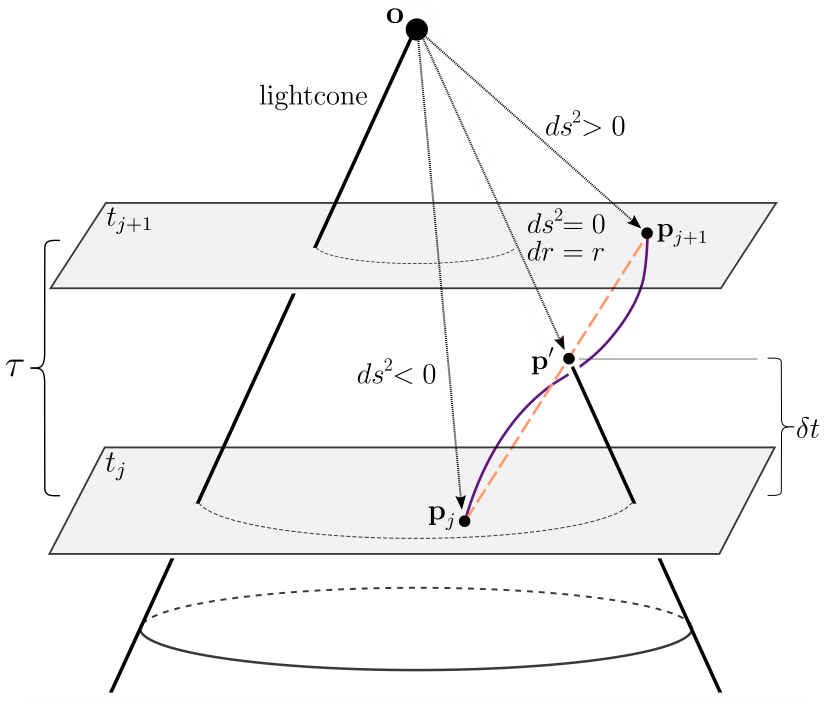}
  \end{minipage}\hspace{0.03\linewidth}
  \begin{minipage}[c]{0.30\linewidth}
    \caption{
    Schematic showing the interpolation process which fills the cosmoDC2 particle lightcone. Each plane represents a projected simulation snapshot, and time increases vertically, with the observer located at \textbf{o}. A particle (events \textbf{p}) is seen crossing the observer's lightcone between snapshots $j$ and $j+1$ along its "true" worldline (unknown), in purple. Interpolation (orange dashed line) is used to solve for $\delta t$ and $r$ via Eq.~\ref{eq:dt_solver}. Event $\textbf{p}'$ is the final output.} \label{fig:particle-interp}
  \end{minipage}
  \hspace{0.1\linewidth}
\end{figure}

Our approach, then, is to linearly interpolate particle positions between the two simulation snapshots that bound the event where the particle's separation $ds^2=0$. For a particular particle, we denote those two bounding snapshots as $j$ and $j+1$. Noting that $t_{j} \leq t_e < t_{j+1}$, we define two useful time quantities,
\begin{align}\label{eq:lc_dt}
\delta t &\equiv t_e - t_j\\
\tau &\equiv t_{j+1} - t_j,
\end{align}
and approximate $r$ via a linear interpolation, as
\begin{equation}\label{eq:lc_rapprox}
r \approx \|\textbf{r}_j + \textbf{v}_\text{lin}\delta t\|,
\end{equation}
where bold-faced quantities are 3-component vectors in comoving Cartesian space, and the ``equivalent linear velocity'' $\textbf{v}_\text{lin}$ is defined as $\textbf{v}_\text{lin} \equiv (\textbf{r}_{j+1} - \textbf{r}_j)/\tau$. In rewriting Eq.~\ref{eq:lc_rw} in terms of Eq.~\ref{eq:lc_dt}-~\ref{eq:lc_rapprox}, we can solve for the unknown $\delta t$. First, we break the time integral on the LHS of Eq.~\ref{eq:lc_rw} into two pieces, one over $[t_e, t_{j+1}]$, and the other over $[t_{j+1}, t_0]$:
\begin{equation}\label{eq:broken_int}
    \int_{t_e}^{t_0}c\frac{dt}{a(t)} = \int_{t_e}^{t_{j+1}}c\frac{dt}{a(t)} + \int_{t_{j+1}}^{t_0}c\frac{dt}{a(t)} .
\end{equation}
The bounds and integrand of the latter piece are entirely known in the simulation parameters and snapshot information, so we evaluate it numerically via Simpson's rule quadrature, and will refer to that result as
$\Theta_{j+1}$. We cannot perform a trivial numerical evaluation for the $[t_e, t_{j+1}]$ piece, however, since the lower bound $t_e$ is unknown. Instead, this integral is simplified and solved analytically; we change the variable of integration from $t$ to $t'=(t-t_j)$, and approximate the result by dropping higher order (>2) terms in $\delta t$ and derivatives in $a$:
\begin{align}\label{eq:int_approx}
    \int_{t_e}^{t_{j+1}}c\frac{dt}{a(t)} &\approx \int_{\delta t}^\tau c\dfrac{dt'}{a_j + \dot{a}_j(t')} \nonumber\\
    &\approx \frac{c}{a_j}\left[(\tau-\delta t) - \frac{\dot{a}_j}{a_j}\frac{(\tau^2-\delta t^2)}{2}\right].
\end{align}
Making the substitutions into Eq.~\ref{eq:lc_rw}, we have
\begin{align}\label{eq:lc_solver}
    \frac{c}{a_j}\left[(\tau-\delta t) \frac{\dot{a}_j}{a_j}\frac{(\tau^2-\delta t^2)}{2}\right]
    + \Theta_{j+1} &= \|\textbf{r}_j\| + \delta t\frac{(\textbf{r}_j \cdot \textbf{v}_\text{lin})}{\|\textbf{r}_j\|},
\end{align}
where, in the final step, we have isolated $\delta t$ in the RHS by applying the binomial series and dropping higher order terms (we print only the first order contribution here for brevity, though the implementation retains up to second order terms in $\delta t$). Finally, Eq.~\ref{eq:lc_solver} can be used to solve for $\delta t$ through a quadratic formula:
\begin{align}\label{eq:dt_solver}
\delta t &= \dfrac{-\beta\pm\sqrt{\beta^2-4\alpha\gamma}}{2\alpha}
\end{align}
where we have
\begin{align}\label{eq:quad_terms}
\alpha &= \dfrac{c\dot{a}_j}{2a_j^2}\nonumber\\
\beta &= -\dfrac{c}{a_j} - \dfrac{(\textbf{r}_j\cdot\textbf{v}_\text{lin})}{\|\textbf{r}_j\|}\\
\gamma &= \dfrac{c\tau}{a_j} - \dfrac{c\dot{a}_j\tau^2}{2a_j^2} - \|\textbf{r}_j\| + \Theta_{j+1}\nonumber
\end{align}

Finally, we can solve for the cosmic time $t_e = t_j+\delta t$ and comoving position $r = \| \textbf{r}_j + \textbf{v}_\text{lin}\delta t\|$, using second-order approximations, for each simulation particle's lightcone crossing. The relevant quantities are illustrated in Figure~\ref{fig:particle-interp}. After doing this for each particle in the simulation box at some snapshot $j$, we write out $r$ and the corresponding scale factor $a_e$ for all particles for which the result of Eq.~\ref{eq:dt_solver} satisfies $0 \leq \delta t < \tau$. The totality of those particles constitute the ``lightcone shell'' for the simulation output at timestep $j$. 

The performance of this routine as applied to the Outer Rim snapshots for cosmoDC2 production is summarized in Sec.~\ref{sec:plc} and Fig.~\ref{fig:lc_pk}. More fine-grained details, as well as an exploration into the inaccuracies introduced to the solver by the approximations made in Eq.~\ref{eq:lc_rapprox}-~\ref{eq:quad_terms}, are given in an extended set of pedagogical notes of our implementation \citep{2019arXiv190608355H}.

For halo lightcones, the process is largely identical, except that the direction of the interpolation is reversed in time, going toward higher redshift-- this is simply because it is a single halo at time $t_{j+1}$ whose properties we want to be represented on the lightcone, rather than a time-averaged description of its progenitors (see Section~\ref{sec:hlc} and Fig.~\ref{fig:halo-interp}). For a given halo at snapshot $j+1$, we compute $\textbf{v}_\text{lin}$ in Eq.~\ref{eq:lc_rapprox}, and perform the interpolation, by taking $r_{j+1}$ to be the position of the halo's potential minimum, and $r_j$ to be the potential minimum of its mose massive progenitor. After applying this technique to all nodes of the FOF merger tree at snapshot $j+1$, a final cleaning step is done to remove all but the most massive fragment of each splitting halo (see \citet{rangel2017building}) from the lightcone dataset.

\pagebreak
\section{Galaxy Properties}
\label{sec:galaxy_app}

\begin{longtable}[h]{ |p{6cm}||p{3cm}|p{5cm}|  }
\caption{This table enumerates the main properties available in the cosmoDC2. Some of properties are defined in the GCR schema (Section~\ref{sec:reader}) while other properties are available only as ``native quantities''. Many properties have variant representations which are labeled with superscripts. For example, most of the luminosity properties can be broken down into a disk, bulge and total component. The full list of variants are listed at the bottom.}   \vspace{0.5em}
\label{tab:catalogcontents}

\\ \hline
\cellcolor{gray!25} Spatial  &  Units & Explanation\\
\hline
\endfirsthead
\multicolumn{3}{l}%
{\tablename\ \thetable\ -- \textit{Continued from previous page}} \vspace{0.5em}
\endhead

\multicolumn{3}{r}{\textit{Continued on next page}} \\
\endfoot
\endlastfoot

 Comoving position  & comoving $h^{-1}$Mpc    & x, y, z co-ordinates\\
 Velocity           & km/s     & x, y, z, radial, total \\
 Projected coordinates &   degrees  & RA/Dec (J2000)  \\
 Cosmological redshift &unitless & based on position only \\
 Total redshift    &unitless & peculiar velocity corrections\\
 \hline
 
 \multicolumn{3}{c}{}\\
 
 \hline
 \cellcolor{gray!25}  Luminosity   &  Units&Explanation\\
 \hline                           
 Broadband filters$^{O/R,~D/B/T,~dust}$  & AB Magnitudes & LSST and SDSS filters, B/V band\\
 SED filter luminosities$^{D/B/T,~dust}$ &   flux  & 30 tophats from 100 nm to  2000 nm \\
 Line luminosities$^{D/B/T}$ & flux & H$\alpha$, H$\beta$, \NII, \OII, \OIII, \SII \\
 Continuum luminosity$^{D/B/T}$ & flux & H, O, Lyc \\
 Luminosity rescaling & unitless & See Section~\ref{sec:lum_adjustment}\\
 \hline
 
 \multicolumn{3}{c}{}\\
 
 \hline
 \cellcolor{gray!25}  Host Halo &  Units&Explanation\\
 \hline
 Unique Halo ID & - &  \\
 Halo mass & $h^{-1}$M$_\odot$    & FoF halo mass, $b=0.168$    \\   
 Halo position & comoving $h^{-1}$Mpc & x, y, z FoF halo potential center \\
 Halo velocity & km/s & x, y, z mean halo velocity\\
 Halo centric galaxy position & comoving $h^{-1}$Mpc & x, y, z galaxy position relative to halo center \\      
 %% Halo tag  &      & \\

 %% MD Mpeak & & \\
 %% MD Mvir & & \\
 %% source/target id/mass/pos & & \\
 % OR Host Halo Spin & & \\ OR halo spin isn't related to the galaxies
 % at all. This will only serve to confuse people. 
 \hline
 
 \multicolumn{3}{c}{}\\
 
 \hline
 \cellcolor{gray!25} Shape &  Units&Explanation\\
 \hline
 Ellipticity components$^{D/B/T}$  & unitless    &$e_1$, $e_2$, $e_{total}$, unlensed\\
 Half Light Radius$^{D/B/T}$       & proper kpc and Arcsec &             \\
 Position angle              & degrees & 0-180, random \\
 Sersic Index$^{D/B}$                & unitless & Combined not provided\\
 \hline

 \multicolumn{3}{c}{}\\
 
 \hline
 \cellcolor{gray!25} Weak Lensing &  Units&Explanation\\
 \hline
 Lensed projected coordinates & degrees & RA/Dec (J2000)\\
 Shear$^{P/T}$ & unitless & $\gamma_1$, $\gamma_2$\\
 Magnification & unitless & \\
 Convergence   & unitless & \\
 \hline

 \pagebreak
 \multicolumn{3}{c}{}\\
 
 \hline
 \cellcolor{gray!25} Galaxy Matter &  Units&Explanation\\
 \hline
 Stellar Mass$^{D/B/T}$ & M$_\odot$ & \\
 SFR$^{D/B/T}$ & M$_\odot$/Gyr& \\
 Stellar Metal Mass$^{D/B/T}$ & M$_\odot$& \\
 Black hole Mass & M$_\odot$ & \\
 Black hole Accretion rate & M$_\odot$/Myr & \\
 Black hole eddington ratio & unitless & \\
 %% Velocity & & ??\\
 \hline

 \multicolumn{3}{c}{}\\
 \hline
 \cellcolor{gray!25} Galaxy Identifiers &  Units&Explanation\\
 \hline
 Unique galaxy ID & -  & \\
 Central     & boolean  & Central or satellite galaxy\\
 Red sequence & boolean & Galaxy modeled as red sequence \\
 %% Galacticus Galaxy ID & & \\
 \hline
 \multicolumn{3}{l}{}\\
 \multicolumn{3}{l}{$^{O/R}$: observer and rest frame }\\
 \multicolumn{3}{l}{$^{D/B/T}$: disk, bulge and total components }\\
 \multicolumn{3}{l}{$^{D/B}$: disk and bulge  components }\\
 \multicolumn{3}{l}{$^{dust}$: with and without host galaxy dust extinction}\\
   \multicolumn{3}{l}{$^{P/T}$: PhoSim and TreeCorr conventions. The PhoSim (TreeCorr) convention is defined with a negative (positive) $\gamma_2$}\\
    \multicolumn{3}{l}{\hspace{0.6cm}value when the major axis of a galaxy is oriented in the NW-SE direction when looking outwards from the Earth}\\
 %% \multicolumn{3}{l}{$^1}$: observer and rest frame }\\
 %% \multicolumn{3}{l}{$^2$: disk, bulge and total components }\\
 %% \multicolumn{3}{l}{$^2$: disk and bulge  components }\\
 %% \multicolumn{3}{l}{$^3$: with and with out host galaxy dust extinction}\\
\end{longtable}

%% \section{...}

\section{The Catalog Reader}
\label{sec:reader}

The LSST DESC uses a number of catalogs to enable scientific analyses and has developed a Python package,
\texttt{GCRCatalogs}\footnote{\url{https://github.com/LSSTDESC/gcr-catalogs} [github.com]}, to facilitate access to these catalogs and to ensure proper version control. The cosmoDC2 galaxy catalog is also released to the LSST DESC via the \texttt{GCRCatalogs} package. In particular, we provide a Python class that implements the ingestion of the catalog data and the translation of internal (native) quantity names and units to a predefined schema. The  \texttt{GCRCatalogs} package obviates the need for
the end users to learn the internal catalog structure or the naming scheme. The package also provides queries that can be used to obtain further information about catalog quantities.

\texttt{GCRCatalogs} uses the Generic Catalog Reader\footnote{\url{https://github.com/yymao/generic-catalog-reader} [github.com]} (GCR)
base class to provide additional features such as data filtering. The raw data format of cosmoDC2 is a set of files based on subvolumes defined by the HEALPix
pixelization scheme and three redshift ranges. With the GCR, users can easily select a subset of data for further use. The \texttt{GCRCatalogs} package
also interfaces with the DESCQA validation framework and the image simulation pipeline (for the generation of the so-called ``instance catalogs'' containing the information for a single exposure). 

Tutorials with example jupyter notebooks showing many use cases for interacting with the cosmoDC2 catalog are available online.\footnote{\url{https://github.com/LSSTDESC/DC2-analysis/tree/master/tutorials}[github.com]} These examples assume that the catalog user has a NERSC account. For those users who download the catalog from the public website, a tutorial is showing how to use the GCR to read a catalog from a local file is available at \url{https://portal.nersc.gov/project/lsst/cosmoDC2/_README.html}.